\documentclass[twocolumn,english,aps,pra,showpacs]{revtex4-1}
\usepackage{mathrsfs}
\usepackage{amsmath}
\usepackage{amssymb}
\usepackage{graphicx}
\usepackage{esint}
\usepackage{babel}
\begin{document}

\title{Effects of initial system-environment correlations on open quantum
system dynamics and state preparation}

\author{Chien-Chang Chen}

\author{Hsi-Sheng Goan}
\email{goan@phys.ntu.edu.tw}
\affiliation{Department of Physics and Center for Theoretical
  Sciences, National Taiwan University, Taipei 10617 }
\affiliation{Center for Quantum Science and Engineering, National Taiwan University, Taipei 10617, Taiwan}

\date{\today}
\begin{abstract}
We investigate the preparation of
 a target initial state for a two-level (qubit) system from a
 system-environment equilibrium 
or correlated  state by an external field.
The system-environment equilibrium or correlated state results from
the inevitable interaction of the system
with its environment. 
An efficient method in an extended
auxiliary Liouville space is introduced 
to describe the dynamics of the non-Markovian open quantum
system in the presence of a strong
field and an initial system-environment correlation.
By using the time evolutions of the population difference, the state
trajectory in the Bloch sphere representation and the trace distance
between two reduced system states of the open quantum system, the
effect of initial 
system-environment correlations on the preparation of a system state
is studied.  
We introduce an upper bound and a lower bound for the trace distance
within our perturbation formalism to describe the diverse behaviors of
the dynamics of the trace distance between
various correlated states after the system state preparation.
These bounds that are much more computable than similar bounds in the
literature give a sufficient condition and a
necessary condition for the increase of the trace distance and are
related to the witnesses of non-Markovianity and initial system-bath
correlation. 

\end{abstract}


\maketitle

\section{INTRODUCTION}

Besides precise coherent control, to be able to prepare initial states
of quantum systems accurately is also one of essential requirements
for quantum information processing. Most of the state preparations
implemented in quantum experiments assume having ideal (closed) quantum
coherent systems. However, almost every quantum system interacts inevitably
with its surrounding environment (bath) resulting in an
equilibrium or correlated system-environment state before any operation
or measurement onto the system is performed \cite{key-01,Breuer02,Rivas12,key-E1,key-E2,key-E3}.
Thus how to prepare a desired initial system state from an equilibrium
or correlated system-environment state becomes an important and practical
issue \cite{Kuah07,Modi11,Gong13A,Gong13B}.

In addition to its effect on the system dynamics, system-environment
correlation also plays a vital role in quantum dynamical maps and
open-system state distinguishability \cite{key-E3,Pechukas94,Royer96,Campisi09,Dijkstra10,key-15,Gong12,key-01-2,key-01-3,key-01-4,key-01-5,key-01-7,Ban09,Uchiyama10,Smirne10,Zhang10,Dajka11,Morozov12,Uchiyama12,Gao13,Semin12,key-14,key-24,key-31,key-30,Rivas14}.
The reduced dynamics of an open quantum system with a tensor product
(factorized) initial
system-environment state is a completely positive map. Completely
positive maps are appealing because they form a time-dependent semigroup
and have a simple mathematical structure that the composition of two
completely positive maps is also a completely positive map \cite{Gorini76,Lindblad76,Breuer02,Rivas12}.
This makes the initial factorization of the joint system-environment 
state an attractive and commonly adopted initial condition when studying
the dynamics of an open quantum system. However, when the
system-environment interaction is not very weak, the initial
system-environment correlation may have an appreciable effect on the
open system dynamics.

References
\cite{Gong13A,Gong13B} 
have investigated the role of initial system-environment correlations
on the reduced system dynamics with a system state initially prepared
by a projective (selective) measurement on the system alone. The
system-environment 
state before the projective measurement is a total thermal equilibrium state
$\rho_{T}^{eq}=e^{-\beta H_{T}}/tr_{T}e^{-\beta H_{T}}$, where $H_{T}$
is the total Hamiltonian, including the system-environment interaction
Hamiltonian. The projective measurement is assumed to be instantaneous
such that the (unnormalized) environment state after the measurement
collapses to $\langle\psi|\rho_{T}^{eq}|\psi\rangle$, conditioned
on the projected system state $|\psi\rangle$. The measurement has
the effect of removing the system-environment correlation, and this
measurement-induced factorized initial system-environment state has
been considered in the literature \cite{Kurizki08,Kurizki09,Kurizki10,Morozov12,Gong13A,Gong13B}.
However, in practice, if the post-measurement system state evolution
is of interest or concern, the measurement made upon the system is
usually nondestructive and indirect and thus takes some finite time
to project the system state to the desired state $|\psi\rangle$.
In other words, unless the measurement is very strong, the system
and environment will evolve away from the total equilibrium state
when the measurement is completed. In this case, the post-measurement
density matrix of the environment is no longer $\langle\psi|\rho_{T}^{eq}|\psi\rangle$.
As a result, the system state preparation by projective measurement
with the above instantaneously measurement-induced factorized system-environment
state is an idealization. We note again that the subsequent system
evolution is dependent on the initial system-environment correlations,
but the initial system-environment state used 
for the investigations
\cite{Kurizki08,Kurizki09,Kurizki10,Morozov12,Gong13A,Gong13B} is,
however, a factorized state. 

In this paper, we investigate an alternative method for state preparation
by applying an external field. 
We use the time evolution of the trace
distance \cite{key-E1,key-E2,key-E3,key-14,key-15,Uchiyama12,key-24}
between two reduced system states of an open quantum system as a
measure of the effect
of initial correlations.  The time evolution
of the trace distance between the quantum states evolving from two kinds
of initial states, a correlated total equilibrium state (or Gibbs
state) and its uncorrelated marginal state for an open qubit system
with a week external driving field was calculated by the rotating-wave
approximation (RWA) in Ref.~\cite{Uchiyama12}.
The open quantum system model considered
in Ref.~\cite{Uchiyama12} before the application of an external
driving field is a pure-dephasing spin-boson model in which the system
operator coupling to the environment commutes with the system Hamiltonian.
However, for the purpose
of accurate state preparations in open quantum systems, fast control
and thus stronger fields beyond the RWA
may be necessary. 

The main purposes of this paper is to investigate under what
conditions the initial factorization approximation of the system-environment
state is valid and how different the system dynamics is when the initial
system-environment correlation is taken into account, especially for
state preparation via applying an external field. We will also
investigate the conditions for the breakdown of the
RWA and the onset of non-RWA corrections for state
preparation.
We go beyond the limitations or approximations imposed in Ref.~\cite{Uchiyama12}
by considering a non-commuting spin-boson model in an external driving
field without making the RWA. The method presented here to derive
the time-nonlocal master equation to second order in the system-bath interaction
and to take the initial correlation into account is based on the
Nakajima-Zwanzig 
projection operator technique \cite{key-12,Breuer02}. To deal with
the application of a strong external field, an efficient formulation
of introducing auxiliary density matrices in an extended Liouville
space to transform the time-nonlocal time-ordered integro-differential
master equation  
into a set of time-local coupled differential equations is employed.
We find that when the driving field strength is above a certain value
the RWA becomes invalid and when the system-environment interaction is
above a certain value, the initial system-environment correlation
become important to the open system dynamics. The detailed values in
relation to the fidelity or error of the state preparation will be
discussed and presented.      

 We will also
investigate the effects of the system-environment correlation
established after 
the state preparation on the subsequent field-free system evolution. 
The dynamics of the
trace distance
has been used to study non-contractivity and non-Markovianity \cite{key-15,key-24,key-28,key-26,Dajka11},
both of which could be induced by initial correlations.
We find that the dynamics of the
trace distances between these correlated states and between
these correlated states and their corresponding factorized states in
the subsequent field-free evolutions exhibit diverse behaviors.
So another purpose of the paper is to introduce an upper
bound and a lower bound for the trace distance within our perturbation
formalism to describe the various behaviors of the dynamics of the
trace distance.
These bounds that we introduce are much more computable than
similar bounds in the literature \cite{key-14,key-24,key-31,key-30}.
These bounds in turn give 
a sufficient condition and a necessary condition for the increase of
the trace distance
and are related to the witnesses of non-Markovianity and initial
system-bath correlations.

The paper is organized as follows. In Sec.~\ref{sec:Initial} we
describe briefly the Hamiltonian of the spin-boson model we study
and the decomposition of the initially correlated state using the
projection operator technique. An efficient method is introduced in
Sec.~\ref{sec:ME} to transform the time-nonlocal master equation,
with an external driving field and initial system-environment correlations
accounted for, to a set of coupled linear time-local equations of
motion in an extended auxiliary Liouville space. Numerical results
and discussions are presented in Sec.~\ref{sec:Results}. In
Sec.~\ref{sec:thermal_state}, we describe how to achieve and express the
total thermal equilibrium 
state as an initially correlated state in the extended Liouville space
formulation. The results to compare the initially correlated state
with the factorized states in terms of time evolutions
of the trace distance between them for state preparation with an external
driving field are presented in Sec.~\ref{sec:state_preparation}. After the
state preparation by applying an external field, the resultant correlated
system-environment states are considered as the initial states for subsequent
field-free evolutions. The diverse behaviors of the dynamics of the
trace distance between these correlated initial states in the subsequent
field-free evolutions are shown in Sec.~\ref{sec:prepared_states}. The
upper and lower bounds of the trace distance similar to those in Refs.~\cite{key-14,key-24,key-31,key-30}
are introduced in Sec.~\ref{sec:bounds} to analyze these
behaviors.  Finally, a 
conclusion is presented in Sec.~\ref{sec:Conclusion}.

\section{INITIAL SYSTEM-ENVIRONMENT CORRELATION}

\label{sec:Initial}

The total Hamiltonian of the dissipative spin-boson model considered
here to study the initial correlations is:
\begin{equation}
H_{T}\left(t\right)=H_{s}\left(t\right)+H_{b}+H_{sb},\label{eq:HT}
\end{equation}
with the two-level (qubit) system Hamiltonian $H_{s}(t)$, the bath (environment)
Hamiltonian $H_{b}$ and the system-bath interaction Hamiltonian $H_{sb}$
given respectively by $\left(\hbar=1\right)$
\begin{eqnarray}
H_{s}\left(t\right) & = & \Omega\sigma_{z}+H_{d}\left(t\right),\label{eq:Hs}\\
H_{b} & = & \sum_{i}\omega_{i}b_{i}^{\dagger}b_{i},\\
H_{sb} & = & \sigma_{x}B.
\end{eqnarray}
Here $\sigma_{i}$ with $i={x,y,z}$ are the Pauli matrices, $\omega_{i}$
and $b_{i}^{\dagger}\left(b_{i}\right)$ are, respectively, the frequency
and the creation (annihilation) operator of the bath mode $i$. The
bath operator $B$ in the the system-bath interaction is $B=\sum_{i}g_{i}\left(b_{i}^{\dagger}+b_{i}\right)$
with the coupling constant $g_{i}$ for the respective bath mode $i$.
The Hamiltonian $H_{d}\left(t\right)$ accounts for the Hamiltonian
of the applied time-dependent external driving field, 
\begin{equation}
H_{d}(t)=\Omega_{R}\cos\left(\omega_{L}t\right)\sigma_{x},\label{eq:Hd}
\end{equation}
where $\Omega_{R}$ is the field strength, also called the Rabi frequency,
and $\omega_{L}$ is the field frequency.

Due to unavoidable interaction between a quantum system and its surrounding
bath, the initial state considered at time $t=0$ before performing
any operation is the correlated total thermal equilibrium state, 
\begin{equation}
\rho_{T}^{eq}=e^{-\beta H_{T}}/tr_{T}e^{-\beta H_{T}},\label{eq:ES_T}
\end{equation}
where $H_{T}$ is the total Hamiltonian without the driving Hamiltonian
$H_{d}(t)$.

The projection operator technique of Nakajima-Zwanzig \cite{Breuer02}
will be used to derive the equation of motion of the reduced system
dynamics with the initial system-environment correlation accounted
for. In this formalism, the projection super-operator $P$ acting
on the total system-environment state separates the bath from the system
via $P\rho_{T}(t)=tr_{b}[\rho_{T}(t)]\otimes\rho_{b}=\rho_{s}(t)\otimes\rho_{b}$,
where  $\rho_{b}$ is some
fixed state of the environment which we take as the bath thermal equilibrium
state, i.e., $\rho_{b}=e^{-\beta H_{b}}/tr_{b}e^{-\beta H_{b}}$, and 
we have used the fact that $\rho_{s}(t)=tr_{b}[\rho_{T}(t)]$ is the
density matrix operator of the reduced system obtained by tracing the
total density matrix operator over the bath degrees of freedom.
The projection super-operator $P$ and a complementary super-operator
$Q$ satisfy the following properties: $P+Q=\mathcal{I}$, $P^{2}=P$,
$Q^{2}=Q$ and $PQ=QP=0$, where $\mathcal{I}$ is an identity operator
in the joint total system-environment state space. Any total system-environment
state $\rho_{T}(t)$ can be expressed as 
\begin{eqnarray}
\mathcal{I}\rho_{T}\left(t\right) & = & P\rho_{T}\left(t\right)+Q\rho_{T}\left(t\right)\nonumber \\
 & = & \rho_{s}\left(t\right)\otimes\rho_{b}+Q\rho_{T}\left(t\right).\label{eq:rhoT}
\end{eqnarray}
The bath (environment) thermal equilibrium state $\rho_{b}$ serves
as the bath reference state and is the bath state of the factorized
system-bath states used throughout this paper. Thus the total initial
state can also be expressed as 
\begin{equation}
\rho_{T}\left(0\right)=\rho_{s}\left(0\right)\otimes\rho_{b}+Q\rho_{T}\left(0\right)\label{eq:3}
\end{equation}
that contains a factorized part $\rho_{s}\left(0\right)\otimes\rho_{b}$
and a non-factorized part $Q\rho_{T}\left(0\right)$.

\section{Time-nonlocal Master  Equation}
\label{sec:ME}

Next we present the time-nonlocal master equation that
incorporates the initial system-environment correlation. The equation
of motion for the total density matrix is given by 
\begin{equation}
\dot{\rho}_{T}\left(t\right) 
=-i\left[H_{T}\left(t\right),\rho_{T}\left(t\right)\right].
\label{dot_rho_T}
\end{equation}
Defining the density matrix of the reduced system as
\begin{equation}
\rho_{s}\left(t\right)=tr_{b}\left[\rho_{T}\left(t\right)\right]=tr_{b}\left[\mathcal{G}_{T}\left(t,0\right)\rho_{T}\left(0\right)\right],\label{eq:4}
\end{equation}
one can formally write 
\begin{eqnarray}
\dot{\rho}_{s}\left(t\right) & = & (-i)tr_{b}\left[H_{T}\left(t\right),\rho_{T}\left(t\right)\right]\\
 & \equiv & tr_{b}\left[\mathscr{L}_{T}\left(t\right)\rho_{T}\left(t\right)\right].\label{eq:5}
\end{eqnarray}
Here the propagator superoperator has a general form of 
\begin{equation}
\mathcal{G}_{j}\left(t,t'\right)\equiv T_{+}exp\left[\intop_{t'}^{t}\mathscr{L}_{j}\left(t''\right)dt''\right]\label{eq:G}
\end{equation}
with $T_{+}$ denoting the time-ordering operator necessary to allow
an explicit time-dependent Hamiltonian \cite{key-12,Breuer02}, and
the Liouville superoperator 
\begin{equation}
\mathscr{L}_{j}\left(t\right)A\equiv-i\left[H_{j}\left(t\right),A\right]\label{eq:L}
\end{equation}
defined as commutator between any operator $A$ and $H_{j}\left(t\right)$
with $j=T$ for the present case. Later we will introduce $\mathscr{L}_{s}$
and corresponding $\mathcal{G}_{s}\left(t,t'\right)$ for the system
alone defined as in Eqs.~(\ref{eq:L}) and (\ref{eq:G}) but with
the replacement of Hamiltonian $H_{j}\left(t\right)\to H_{s}\left(t\right)$.
It is then straightforward to verify that 
\begin{equation}
\frac{\partial}{\partial t}\mathcal{G}_{j}\left(t,t'\right)=\mathscr{L}_{j}\left(t\right)\mathcal{G}_{j}\left(t,t'\right).\label{eq:G_EOM}
\end{equation}

After applying the projector operators $P$, $Q$ to Eq. (\ref{dot_rho_T})
and taking terms up to second order in system-bath interaction strength,
one obtains the time-convolution (time-nonlocal) master equation in
the interaction picture \cite{Breuer02}: 
\begin{alignat}{1}
\dot{\tilde{\rho}}_{s}\left(t\right)= & -itr_{b}\left[\tilde{H}_{sb}\left(t\right),Q\rho_{T}\left(0\right)\right]\nonumber \\
 & -\int_{0}^{t}tr_{b}\left[\tilde{H}_{sb}\left(t\right),\left[\tilde{H}_{sb}\left(t'\right),\tilde{\rho}_{s}\left(t'\right)\otimes\rho_{b}\right]\right]dt',\label{eq:6}
\end{alignat}
where $\tilde{\rho}_{s}(t)=\mathcal{G}_{s}\left(0,t\right)\rho_{s}(t)$,
$\tilde{H}_{sb}\left(t\right)=\tilde{\sigma}_{x}\left(0,t\right)B\left(t\right)$,
$\tilde{\sigma}_{x}\left(0,t\right)=\mathcal{G}_{s}\left(0,t\right)\sigma_{x}$,
and $B\left(t\right)=\sum_{i}g_{i}\left(b_{i}^{\dagger}e^{i\omega_{i}t}+b_{i}e^{-i\omega_{i}t}\right)$.
So the non-factorized part $Q\rho_{T}\left(0\right)$ in Eq.~(\ref{eq:3})
now plays a role of an inhomogeneous term in the master equation.
Since there is already an interaction Hamiltonian $\tilde{H}_{sb}(t)$
in the first term on the right-hand side of Eq.~(\ref{eq:6}),
we only need to keep $Q\rho_{T}\left(0\right)$ to first order in
$\tilde{H}_{sb}$. The initial correlated thermal equilibrium state,
Eq.~(\ref{eq:3}), 
to first order in the system-bath interaction Hamiltonian before turning
on the external field is \cite{key-12}: 
\begin{eqnarray}
\rho_{T}\left(0\right) & = & \rho_{T}^{eq}\label{eq:1}\\
 & \simeq & \left(\rho_{s}^{eq}\otimes\rho_{b}\right)\left(1-\int_{0}^{\beta}\tilde{H}_{sb}\left(-i\beta'\right)d\beta'\right),\label{eq:2}
\end{eqnarray}
where $\rho_{s}^{eq}=e^{-\beta H_{s}}/tr_{s}e^{-\beta H_{s}}$ and
$\tilde{H}_{sb}\left(-i\beta'\right)=e^{\beta'\left(H_{s}+H_{b}\right)}H_{sb}e^{-\beta'\left(H_{s}+H_{b}\right)}$.
The first term of Eq. (\ref{eq:2}) is the commonly used initial product
system-environment state and the second term is the first order non-factorized
part $Q\rho_{T}\left(0\right)$ of the total thermal equilibrium state.
Tracing over the bath degrees of freedom in Eq.~(\ref{eq:6}), and
going back to Schr\"odinger picture, we arrive at \cite{key-12,key-20}:
\begin{eqnarray}
\dot{\rho}_{s}\left(t\right) & = & \mathscr{L}_{s}\left(t\right)\rho_{s}\left(t\right)+\mathscr{L}_{x}\left[\mathcal{K}\left(t\right)+\mathcal{K}^{\dagger}\left(t\right)\right],\label{eq:7}\\
\mathcal{K}\left(t\right) & = & -i\int_{-\infty}^{t}C\left(t-t'\right)\mathcal{G}_{s}\left(t,t'\right)\sigma_{x}\rho_{s}\left(t'\right)dt',\label{eq:8}
\end{eqnarray}
where $\rho_{s}\left(t'\leq0\right)=\rho_{s}^{eq}$, $\mathscr{L}_{x}\mathcal{K}\left(t\right)=-i\left[\sigma_{x},\mathcal{K}\left(t\right)\right]$,
and the bath correlation function is given by \cite{key-01,key-12,Breuer02}:
\begin{eqnarray}
C\left(t-t'\right) & \equiv & tr_{b}\left[B\left(t\right)B\left(t'\right)\rho_{b}\right]\nonumber \\
 & = & \int_{0}^{\infty}d\omega J\left(\omega\right)\cos\left[\omega\left(t-t'\right)\right]\coth\left(\frac{\beta\omega}{2}\right)\nonumber \\
 &  & -i\int_{0}^{\infty}d\omega J\left(\omega\right)\sin\left[\omega\left(t-t'\right)\right],\label{eq:9}
\end{eqnarray}
with spectral density $J\left(\omega\right)=\sum_{i}g_{i}^{2}\delta\left(\omega-\omega_{i}\right)$.
The integral from $t=-\infty$ to $t=0$, i.e., $\mathcal{K}\left(0\right)$
in Eq. (\ref{eq:8}), comes from the first term on the right-hand side
of Eq. (\ref{eq:6}) due to the non-factorized contribution
$Q\rho_{T}\left(0\right)$ 
of the initial system-bath state at $t=0$ and indicates that the
inhomogeneous term is the past memory of the homogeneous term in the memory
kernel of the master equation.

To deal with the time-nonlocal time-ordered integro-differential
master equation with a time-dependent 
driving Hamiltonian without making the RWA, 
we express the bath correlation function
as a sum of exponentials \cite{key-12,key-20,key-18,key-22,key-21}
\begin{equation}
C\left(t-t'\right)=\sum_{k}\alpha_{k}e^{\gamma_{k}\left(t-t'\right)}\label{eq:10}
\end{equation}
with complex numbers $\alpha_{k}$ and $\gamma_{k}$ obtained from
numerical methods. By inserting Eq. (\ref{eq:10}) into Eq. (\ref{eq:8}),
one then obtains $\mathcal{K}\left(t\right)=\sum_{k}\mathcal{K}_{k}\left(t\right)$,
where $\mathcal{K}_{k}\left(t\right)=-i\int_{-\infty}^{t}\alpha_{k}e^{\gamma_{k}\left(t-t'\right)}\mathcal{G}_{s}\left(t,t'\right)\sigma_{x}\rho_{s}\left(t'\right)dt'$.
By taking the time derivative of $\mathcal{K}_{k}\left(t\right)$
with the help of the property $\frac{\partial}{\partial t}\mathcal{G}_{s}\left(t,t'\right)=\mathscr{L}_{s}\left(t\right)\mathcal{G}_{s}\left(t,t'\right)$,
Eqs.~(\ref{eq:7}) and (\ref{eq:8}) now become a set of coupled linear
time-local equations: 
\begin{eqnarray}
\dot{\rho}_{s}\left(t\right) & = & \mathscr{L}_{s}\left(t\right)\rho_{s}\left(t\right)+\mathscr{L}_{x}\sum_{k}\left[\mathcal{K}_{k}\left(t\right)+\mathcal{K}_{k}^{\dagger}\left(t\right)\right],\label{eq:11}\\
\dot{\mathcal{K}}_{k}\left(t\right) & = & \left[\mathscr{L}_{s}\left(t\right)+\gamma_{k}\right]\mathcal{K}_{k}\left(t\right)-i\alpha_{k}\sigma_{x}\rho_{s}\left(t\right).\label{eq:12}
\end{eqnarray}
Similar equations to ${\mathcal{K}_k}(t)$ hold for the Hermitian conjugate 
${\mathcal{K}_k}^\dagger(t)$. 
Compared to solving the reduced density matrix of the system
$\rho_s(t)$ directly through 
Eqs.~(\ref{eq:7}) and (\ref{eq:8}), the resultant coupled differential equations for 
$\{\rho_s(t),\mathcal{K}_k,{\mathcal{K}_k}^\dagger; k=1,2,3,...\}$ 
are easy to solve as they are time-local and free from the
time-ordering and memory kernel integration problems. 
The factorized part of the initial condition at $t=0$ 
is $\rho_{s}\left(0\right)$, and the non-factorized part of the initial
condition is $\mathcal{K}_{k}\left(0\right)$ 
and $\mathcal{K}_{k}^{\dagger}\left(0\right)$.
If $Q\rho_{T}\left(0\right)=0$, i.e., the initial system-bath state
is in fact a factorized state, $\mathcal{K}_{k}\left(0\right)=\mathcal{K}_{k}^{\dagger}\left(0\right)=0$.

\section{NUMERICAL RESULTS}
\label{sec:Results}

We use the trace distance, a measure of distance between two quantum states,
to quantify and discuss our results. Formally, the trace distance is defined
as distinguishability between two states $\rho^{\alpha}$, $\rho^{\beta}$
by the expression \cite{key-26,Dajka11}: 
\begin{equation}
{D}\left(\rho^{\alpha},\rho^{\beta}\right)=\frac{\left\Vert \rho^{\alpha}-\rho^{\beta}\right\Vert }{2}=\frac{1}{2}tr\sqrt{\left(\rho^{\alpha}-\rho^{\beta}\right)^{2}}.\label{eq:TD}
\end{equation}
It can be shown that the trace distance between two reduced system
states $\rho_{s}^{\alpha}\left(t\right)$ and
$\rho_{s}^{\beta}\left(t\right)$
of a  qubit system
at time $t$ in 2-dimensional Hilbert space has the following form:
\cite{key-15}: 
\begin{eqnarray}
{D}\left(\rho_{s}^{\alpha}\left(t\right),\rho_{s}^{\beta}\left(t\right)\right) & = & \frac{1}{2}\left\Vert \Delta\rho_{s}^{\alpha,\beta}\left(t\right)\right\Vert =\frac{1}{2}tr\sqrt{\left(\Delta\rho_{s}^{\alpha,\beta}\left(t\right)\right)^{2}}\nonumber \\
 & = & \sqrt{\left(\Delta\rho_{11}^{\alpha,\beta}\left(t\right)\right)^{2}+\left|\Delta\rho_{12}^{\alpha,\beta}\left(t\right)\right|^{2}},
\label{eq:D(rho_alpha_beta)}
\end{eqnarray}
where $\Delta\rho_{j}^{\alpha,\beta}\left(t\right)=\rho_{j}^{\alpha}\left(t\right)-\rho_{j}^{\beta}\left(t\right)$
with subscript $j=s$ for the reduced system density matrix and $j=11,12$
for the matrix elements of $\rho_{s}\left(t\right)=\begin{pmatrix}\rho_{11}\left(t\right) & \rho_{12}\left(t\right)\\
\rho_{12}^{*}\left(t\right) & 1-\rho_{11}\left(t\right)
\end{pmatrix}$.

The fidelity between two states $\rho_{s}^{\alpha}$ and $\rho_{s}^{\beta}$ is normally defined by $\mathcal{F}\left(\rho^{\alpha},\rho^{\beta}\right)=\left(tr\sqrt{\sqrt{\rho^{\alpha}}\rho^{\beta}\sqrt{\rho^{\alpha}}}\right)^{2}$.
If $\rho^{\beta}$ is denoted as our prepared state, and $\rho^{\alpha}=\left|1\right\rangle \left\langle 1\right|$
as our target excited state, then the fidelity  can be written as the 
diagonal component of $\rho_{s}^{\beta}$ as
$\mathcal{F}\left(\rho^{\alpha},\rho^{\beta}\right)=\left\langle
  1\right|\rho^{\beta}\left|1\right\rangle=\rho_{11}^{\beta}$. 
The error may be defined to be
$1-\mathcal{F}\left(\rho^{\alpha},\rho^{\beta}\right)=1-\rho_{11}^{\beta}$;
but this definition discards the difference in off-diagonal terms
between the prepared state and the target state. 
If one instead uses the trace distance as the
definition for error
between the two states $\rho_{s}^{\alpha}=\left|1\right\rangle
\left\langle 1\right|$ and $\rho_{s}^{\beta}$ (degrees of deviation of the
prepared state from the target excited state), 
then one obtains from Eq.~(\ref{eq:D(rho_alpha_beta)}) 
\begin{equation}
{D}\left(\rho^{\alpha},\rho^{\beta}\right)=\sqrt{\left(1-\rho_{11}^{\beta}\right)^{2}
 +\left|\rho_{12}^{\beta}\right|^{2}}.  
\label{eq:error_ES}
\end{equation}
One can see that the error definition by the trace distance,
Eq.~(\ref{eq:error_ES}), takes 
the differences in both the diagonal and off-diagonal components into account. 
If the off-diagonal term $\rho_{12}^{\beta}$ is ignored, then the
trace distance reduces to the error definition of
$1-\mathcal{F}\left(\rho^{\alpha},\rho^{\beta}\right)$. 
Thus in the following, we use Eq.~(\ref{eq:error_ES})
as the definition of error 
for the state preparation.

In principle, we could deal with any given form of the spectral density.
But as a particular example, we consider an ohmic spectral density
\begin{equation}
J\left(\omega\right)=\frac{\xi}{2}\omega e^{-\omega/\omega_{c}},\label{eq:SD}
\end{equation}
where $\xi$ is a dimensionless system-bath coupling constant, and
$\omega_{c}$ is the bath cutoff frequency. The bath correlation function,
Eq.~(\ref{eq:9}), is numerically fitted in a multi-exponential form 
as Eq.~(\ref{eq:10}).
The minimal number of fitting exponential terms $k$ is chosen to
let the value of the squared 2-norm of the residual between Eq. (\ref{eq:9})
and Eq. (\ref{eq:10}) be less than or equal to 10$^{-7}$. Only three
to six terms in the expansion are required to fit with CPU time from
seconds to minutes regardless of the bath temperature. This is in
contrast to the method of the spectral density parametrization
\cite{key-12,key-20,key-18,key-22} 
which requires more than 48 exponential terms to express the same
bath correlation function at a low temperature of $k_{B}T=1/\beta=0.2\Omega$
\cite{key-12}. Through 
out this paper, the cutoff frequency of $\omega_{c}=7.5\Omega$ for the
bath spectral density is used and
a lower temperature of $k_B T =1/\beta= 0.1\Omega$  is chosen
such that a higher state preparation fidelity or smaller error 
can be achieved.

\begin{figure}
\includegraphics[width=1\columnwidth]{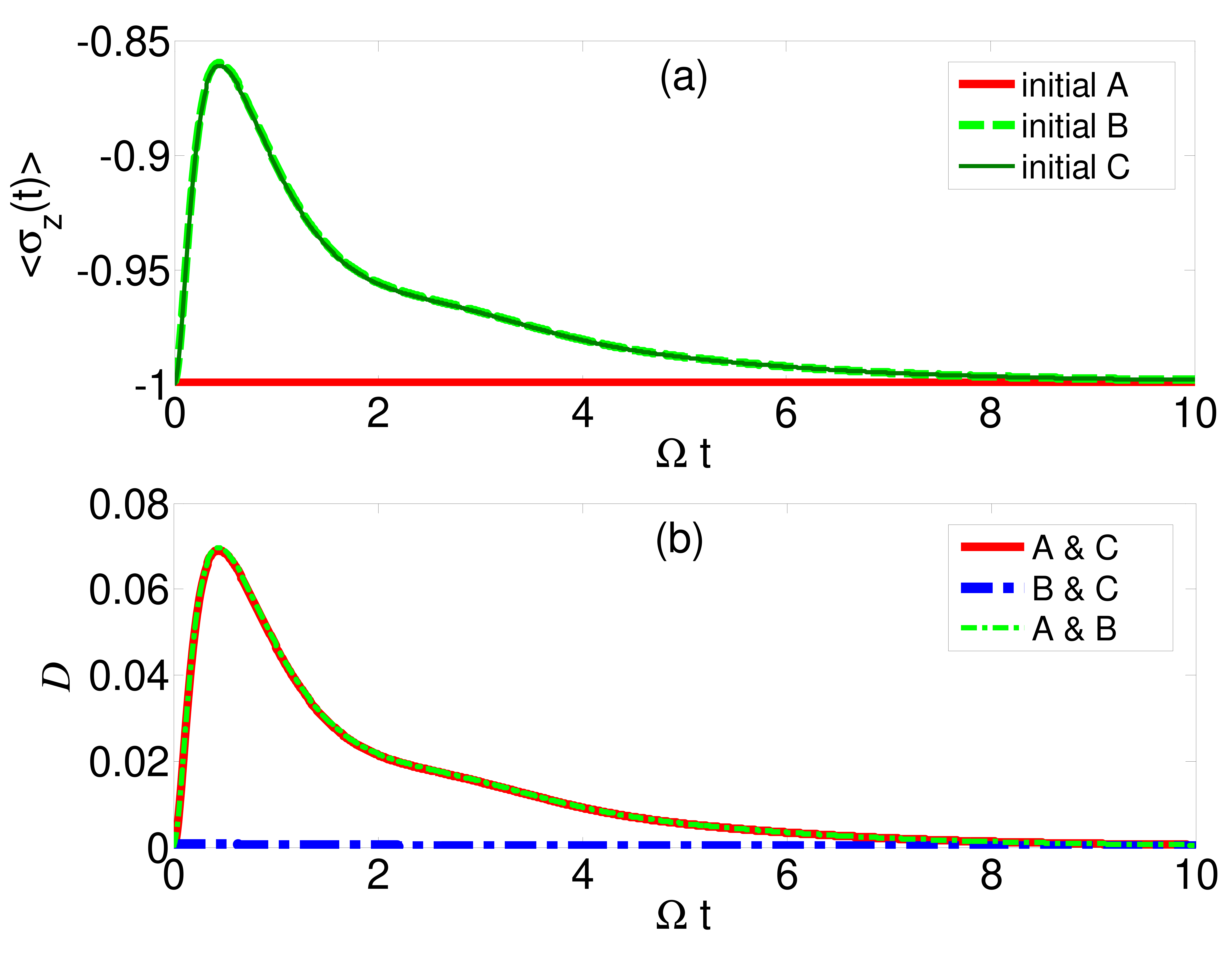} 
\caption{(Color online) (a) Field-free time evolutions of $\left\langle \sigma_{z}\left(t\right)\right\rangle =tr_{s}\left[\sigma_{z}\rho_{s}\left(t\right)\right]$
for different initial states of Initial-A: $\rho_{T}^{eq}$, Initial-B:
$tr_{b}\left(\rho_{T}^{eq}\right)\otimes\rho_{b}$ and Initial-C:
$\rho_{s}^{eq}\otimes\rho_{b}$, the first term of Eq. (\ref{eq:2}).
(b) Trace distance between the reduced system states evolving from
the different initial states in (a). The line ``A \& B'' denotes
the evolution of the trace distance ${D}\left(\rho_{s}^{A}\left(t\right),\rho_{s}^{B}\left(t\right)\right)$
between the two reduced system states $\rho_{s}^{A}\left(t\right)$
and $\rho_{s}^{B}\left(t\right)$ evolving, respectively, from the
two initial states Initial-A and Initial-B as shown in (a), similarly
for lines ``A \& C'' and ``B \& C''. Other parameters are $\xi=0.1$,
$\omega_{c}=7.5\Omega$ and $k_{B}T=1/\beta=0.1\Omega$.}
\label{fig:1} 
\end{figure}

\subsection{Initial state preparations to system-bath thermal  equilibrium state}
\label{sec:thermal_state}

The correlated thermal equilibrium state $\rho_{T}^{eq}$ of Eq. (\ref{eq:ES_T})
that will be used as a correlated initial state for Eqs. (\ref{eq:11})
and (\ref{eq:12}) up to second order can be obtained by different
methods. 
One method is to calculate $\rho_{s}\left(0\right)$ by directly
performing the second-order expansion of Eq.~(\ref{eq:ES_T}) 
and to calculate $\mathcal{K}_{k}\left(0\right)$ 
by performing the integral of Eq. (\ref{eq:8}).
An alternative method that is simpler
is to numerically propagate Eqs.~(\ref{eq:11}) and (\ref{eq:12})
without the application of any external field from any (factorized)
state, say, the first term of Eq. (\ref{eq:2}), for sufficiently
long time to reach equilibrium. Then the reduced system state of the
total equilibrium state is $tr_{b}\left(\rho_{T}^{eq}\right)=\rho_{s}\left(t\rightarrow\infty\right)$
and the corresponding auxiliary equilibrium density matrices
are $\mathcal{K}_{k}^{eq}=\mathcal{K}_{k}\left(t\rightarrow\infty\right)$.
Here $t\rightarrow\infty$ just means a sufficiently long time at
which the system state no longer changes. Then the obtained 
$tr_{b}\left(\rho_{T}^{eq}\right)$ and $\mathcal{K}_{k}^{eq}$ will be
taken as the initial conditions for the dynamics of the qubit system under
the applications of an external driving field. 
The difference between the
results obtained by these
two methods for the correlated total thermal equilibrium state 
of $tr_{b}\left(\rho_{T}^{eq}\right)$ and $\mathcal{K}_{k}^{eq}$
comes from the fourth-order system-environment coupling
which is negligible here \cite{key-23}. Thus we will adopt the latter
numerical method for the preparation of the correlated total equilibrium
state.

To demonstrate that this numerical method indeed leads to an equilibrium state,
we take $\rho_{s}\left(0\right)=\rho_{s}\left(t\rightarrow\infty\right)$
and non-factorized part of $\mathcal{K}_{k}\left(0\right)=\mathcal{K}_{k}^{eq}$
as the initial conditions for a field-free evolution to compare with
the field-free evolutions of other factorized initial states in which
$\mathcal{K}_{k}\left(0\right)=0$. 
We denote Initial-A as $\rho_{T}^{eq}$ which is the non-factorized
total thermal equilibrium state and is obtained by propagating
Eqs.~(\ref{eq:11}) 
and (\ref{eq:12}) to equilibrium as mentioned above, Initial-B as
$tr_{b}\left(\rho_{T}^{eq}\right)\otimes\rho_{b}$ which is the factorized
part of Initial-A in the decomposition of the projection operator $P$
and has the second-order corrections to the reduced system state
included, and Initial-C as $\rho_{s}^{eq}\otimes\rho_{b}$ which is
a factorized state with the system and the bath being in their respective
individual equilibrium states and is just the first term of Eq.~(\ref{eq:2}).
Figure \ref{fig:1}(a) shows the time evolution of $\left\langle \sigma_{z}\left(t\right)\right\rangle =tr_{s}\left[\sigma_{z}\rho_{s}\left(t\right)\right]$
starting from these initial states. One can see that for the Initial-A
state, $\left\langle \sigma_{z}\left(t\right)\right\rangle $, as
expected, stays the same and does not change at all times, while for
the Initial-B and Initial-C states, it undergoes an appreciable evolution
to different values and eventually reaches the same equilibrium value
as that for Initial-A state. The trace distance between the time-dependent
states of the reduced system with these initial states are presented
in Fig.~\ref{fig:1}(b). Note that the Initial-A and Initial-B states
have the same reduced system density matrix, thus the same initial
value of
$\langle\sigma_{z}\left(0\right)\rangle=tr_{s}\left[\sigma_{z}\rho_{s}\left(0\right)\right]$, 
and a zero initial trace distance. The corresponding time evolution of
the trace distance
between the reduced system states 
in the decomposition of the same initial bath reference state
does increase from its initial value, revealing the evidence of the
initial correlation \cite{key-14,key-30}. The time-dependent trace
distance for the two different initial states then diminishes and
finally returns, in the long-time limit, to zero, indicating that
the reduced system states are the same  due to the fact that 
the corresponding total system-environment
states reach equilibrium and become the same. The nearly zero trace
distance at all times for the case of ``B \& C'' in Fig.~\ref{fig:1}(b),
in addition to the almost identical time evolution $\left\langle \sigma_{z}\left(t\right)\right\rangle $
for the Initial-B and Initial-C states in Fig.~\ref{fig:1}(a), shows
that $tr_{b}\left(\rho_{T}^{eq}\right)=\rho_{s}\left(t\rightarrow\infty\right)$
is almost equal to $\rho_{s}^{eq}$. This is because the first-order
contribution to the system reduced density matrix is zero as
$tr_b[\tilde{H}_{sb}\rho_b]=0$ and 
the second-order
corrections to the system reduce density matrix vanish in the symmetrical
spin-boson model \cite{key-23} which is just the field-free model
we consider. 
Thus in the following discussion, Initial-B and Initial-C
are regarded as the same and only one of them, Initial-C state, is
employed.

\begin{figure}
\includegraphics[width=1\columnwidth]{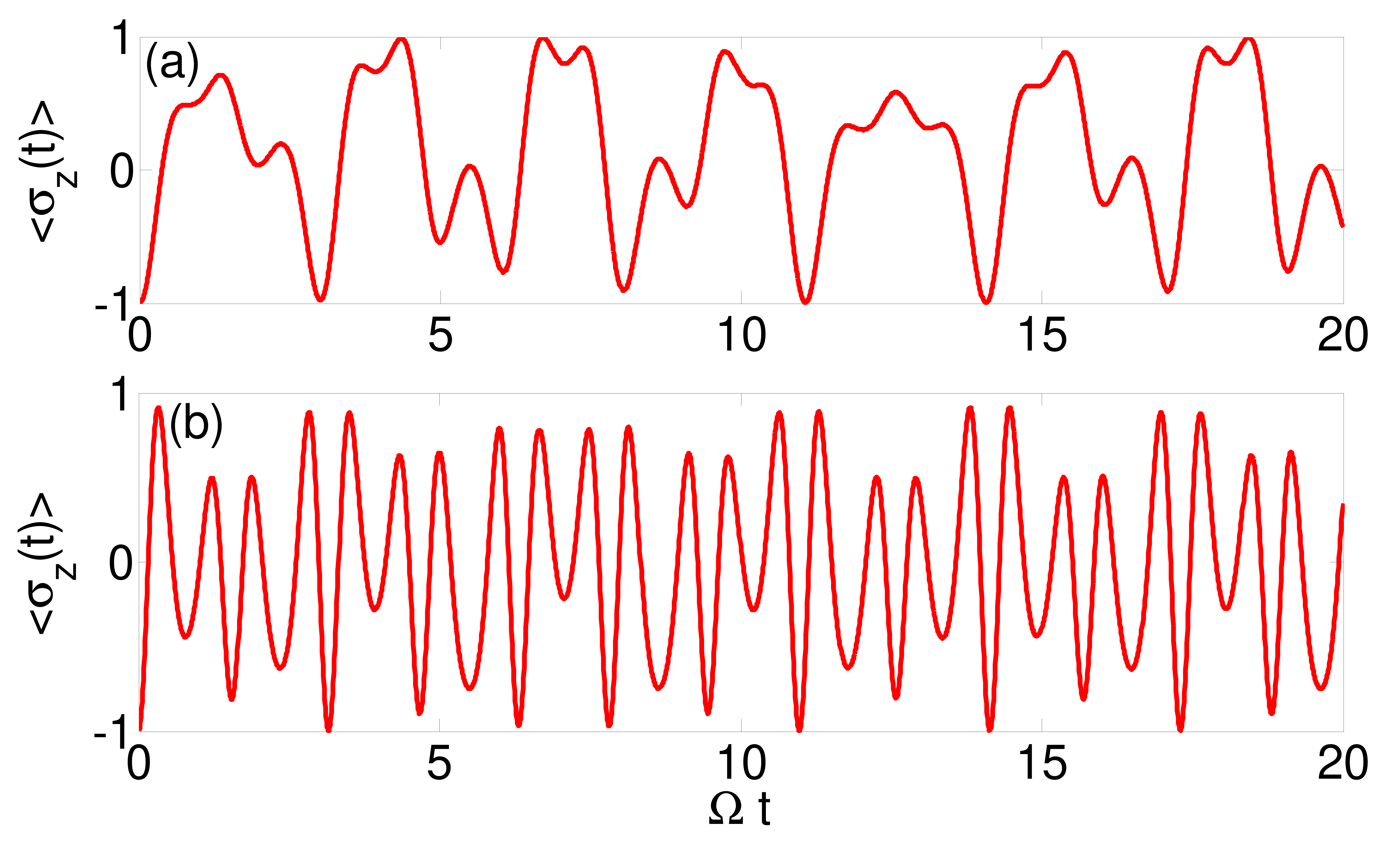}
\caption{(Color online) Unitary time evolutions of $\left\langle
    \sigma_{z}\left(t\right)\right\rangle $ 
of a two-level system initially in the ground state
$\left|0\right\rangle \left\langle 0\right|$ 
driven by an external field with
the Hamiltonian of Eq.~(\ref{eq:Hd}) at resonance 
(i.e., $\omega_{L}=2\Omega$) 
for pulse strengths of (a) $\Omega_{R}=2.4\Omega$ and (b) $\Omega_{R}=5\Omega$.}
\label{fig:2} 
\end{figure}

\begin{figure*}
\includegraphics[width=1\textwidth]{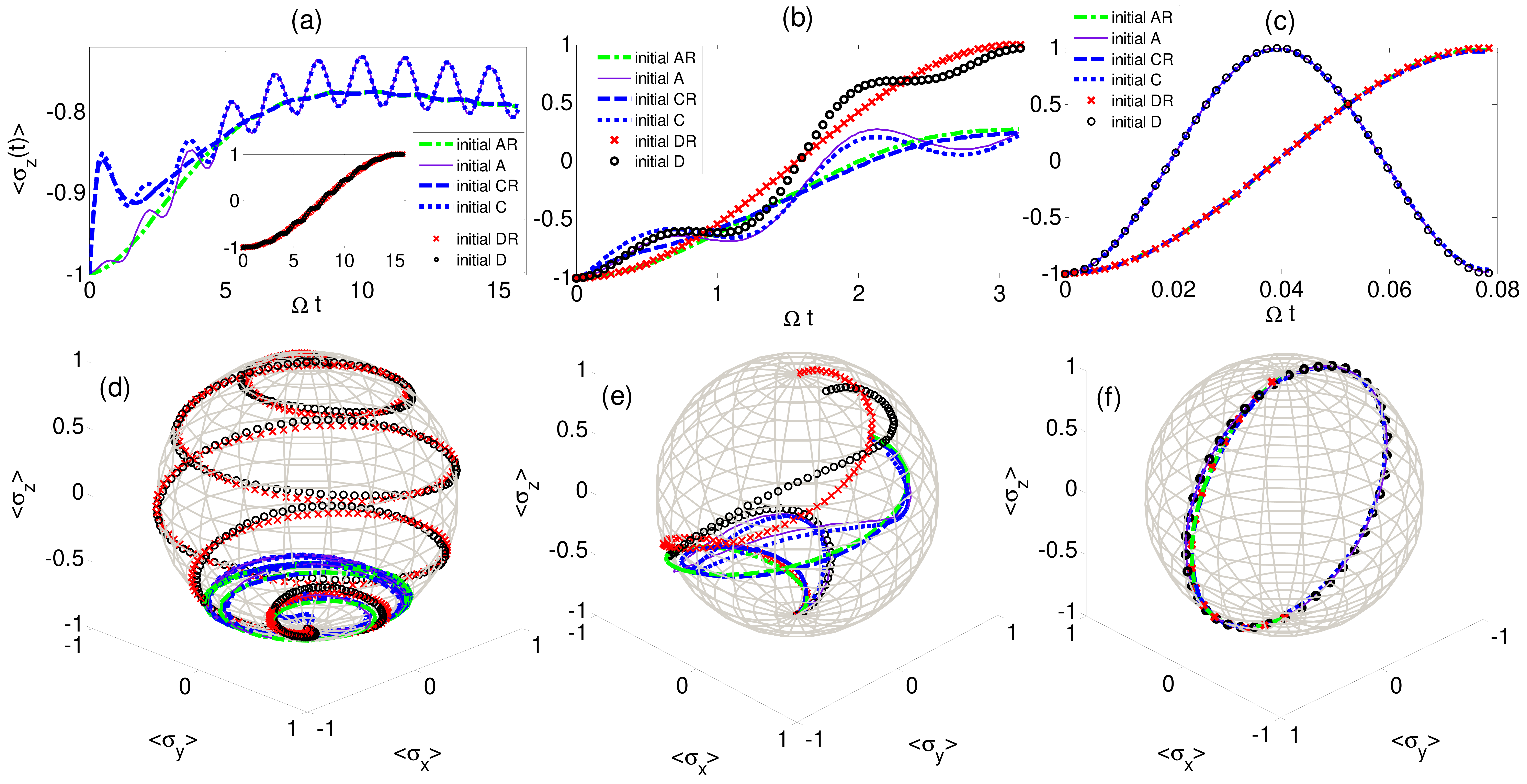}
\caption{(Color online) Time evolutions of 
$\left\langle \sigma_{z}\left(t\right)\right\rangle $
for the excited-state preparations of a two-level system from 
initial states of Initial-A (thin purple solid line),
Initial-C (blue dotted line) and  Initial-D (black circle). 
The Initial-A and Initial-C states are defined in Fig.~\ref{fig:1}. 
The time evolution of 
Initial-D denotes the ideal evolutions
from the ground state 
$\left|0\right\rangle \left\langle 0\right|$ in the absence of the bath.  
The evolutions using the
pulse of the RWA Hamiltonian Eq.~(\ref{eq:15}) from the same
initial states are labeled as Initial-AR (green dot-dashed line), 
Initial-CR (blue dashed line) and Initial-DR (red cross),
respectively. 
The pulse durations are $t=\pi/\Omega_{R}$ for different pulse
strengths of (a) $\Omega_{R}=0.2\Omega$, (b) $\Omega_{R}=\Omega$ and
(c) $\Omega_{R}=40\Omega$. 
The corresponding evolutions of the trajectories in the Bloch sphere
representation for (a), (b) and (c) evolving from the south pole to
the north pole are shown in (d), (e), and (f), respectively. The
evolutions of the states
without the RWA in (f) have about a full-round trajectory while the states with
the RWA have about a half-round trajectory. Other parameters are $\omega_{L}=2\Omega$,
$\xi=0.1$, $\omega_{c}=7.5\Omega$ and $k_{B}T=1/\beta=0.1\Omega$.}
\label{fig:3} 
\end{figure*}

\begin{figure*}
\includegraphics[width=1\textwidth]{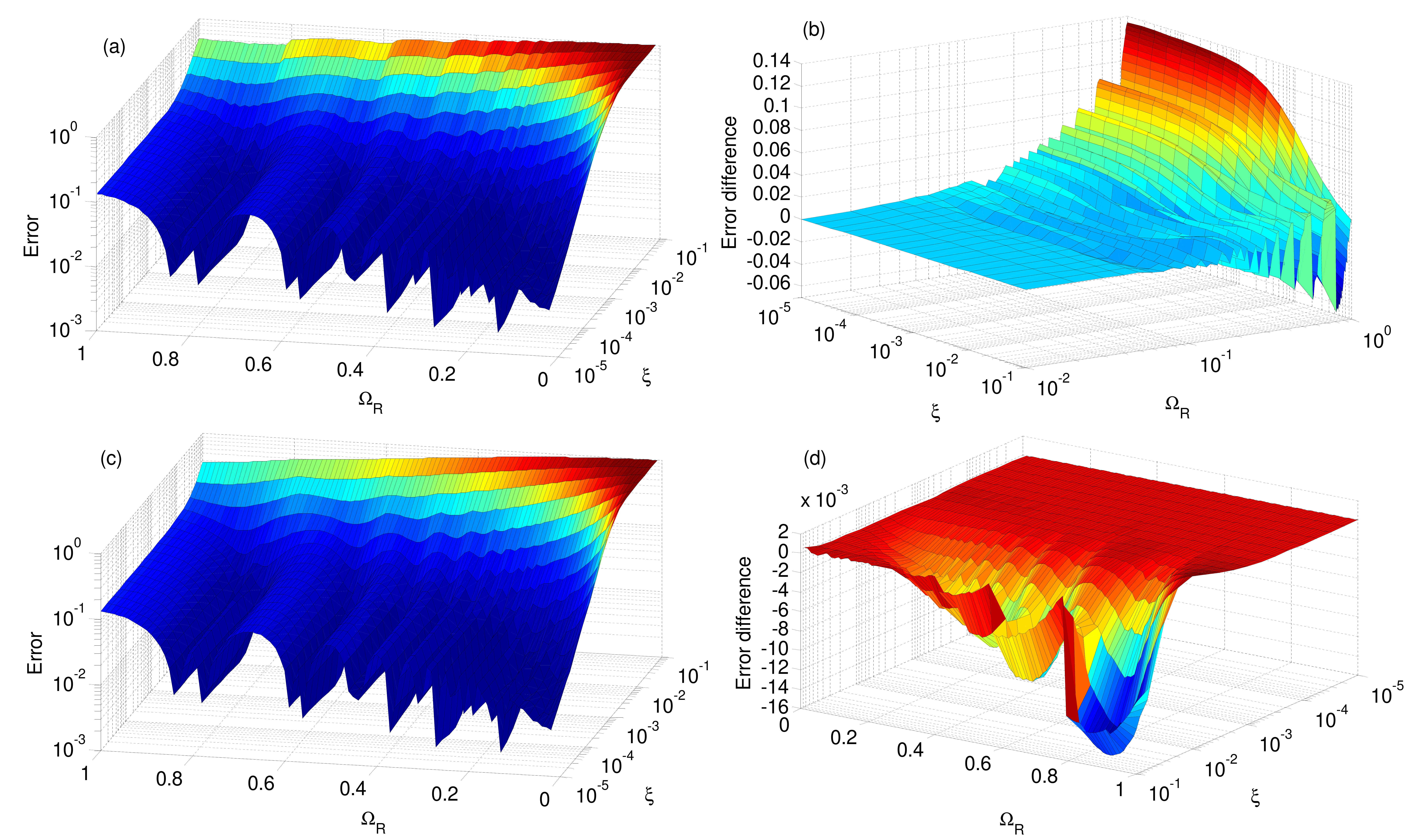} 
\caption{(Color online) Error of the excited state preparation as functions of driving
  strength $\Omega_{R}$ 
and coupling constant $\xi$. (a) Error of the prepared excited states by
the non-RWA pulse using the Hamiltonian of Eq.~(\ref{eq:Hd}) from the
Initial-A state. Pulse duration is determined 
by minimizing the error around the time $t=\pi/\Omega_{R}$ in the
ideal unitary evolution.  
(b) Error difference
between the error in (a) and the one by the $\pi$ pulse of RWA
Hamiltonian of Eq.~(\ref{eq:15}), both starting from the Initial-A
state. (c) Error of the prepared excited states by
the non-RWA pulse with the same setup as in (a) 
but with the pulse duration determined 
by minimizing the error in the open system case.
(d) Error difference between the error in (c) and the
error with the same setup and pulse duration as in (c) but with different
initial state, Initial-C. 
Other parameters are $\omega_{L}=2\Omega$, $\omega_{c}=7.5\Omega$
and $k_{B}T=1/\beta=0.1\Omega$.}
\label{fig:4} 
\end{figure*}

\begin{figure}
\includegraphics[width=1\columnwidth]{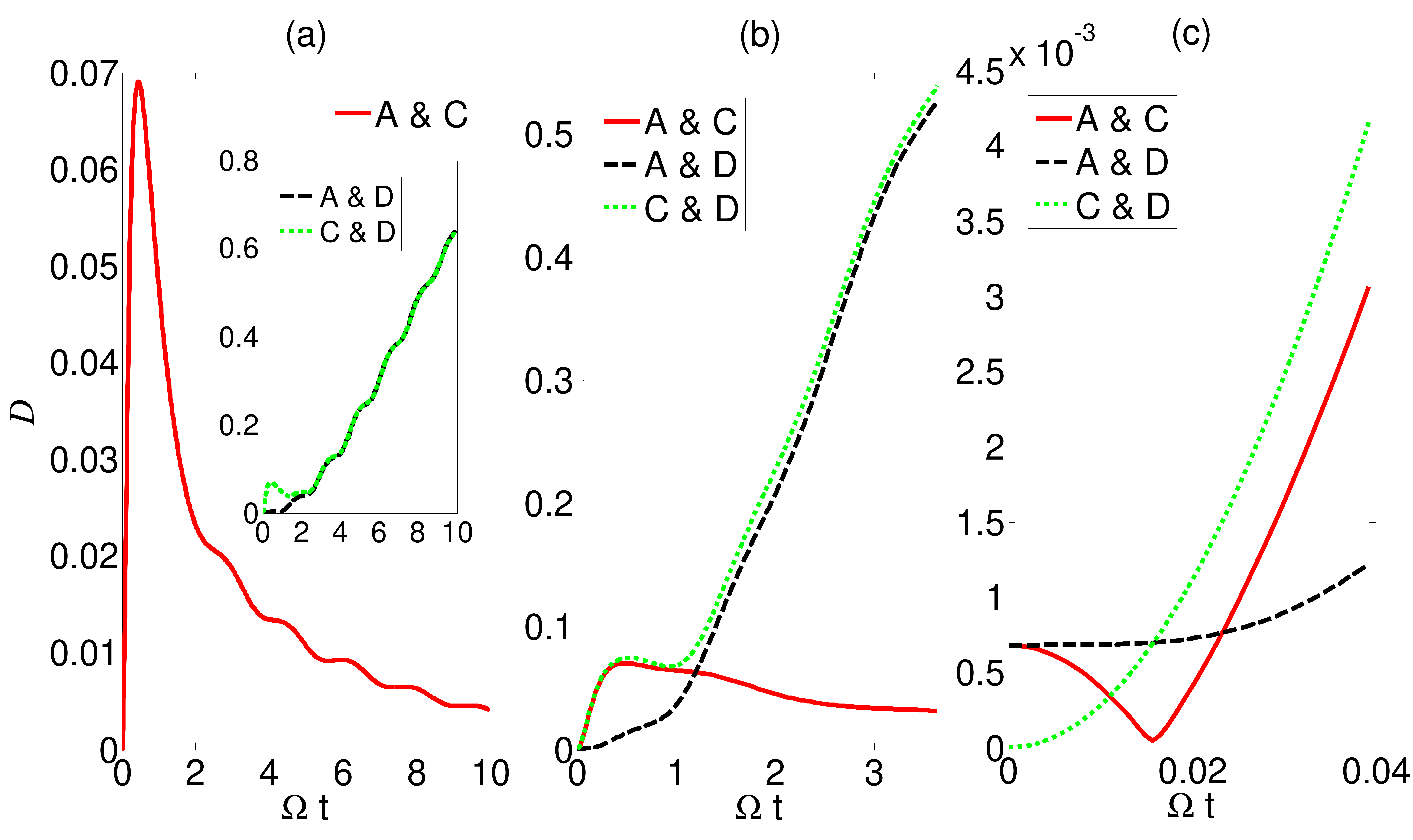}
\caption{(Color online) Time evolutions of the trace distances between the reduced system states evolving from
the Initial-A, initial-C and initial-D states, abbreviated as 
``A \& C'' (red solid line), ``A \& D'' (black dashed line) and 
``C \& D'' (green dotted line), respectively,  
for different pulse strengths of  (a) $\Omega_{R}=0.2\Omega$,
(b) $\Omega_{R}=\Omega$ and (c) $\Omega_{R}=40\Omega$.
The duration of the preparation pulses is obtained by minimizing the error 
for the open system starting from the Initial-A state for each pulse
strength.
Other parameters
are $\omega_{L}=2\Omega$, $\xi=0.1$, $\omega_{c}=7.5\Omega$ and
$k_{B}T=1/\beta=0.1\Omega$.}
\label{fig:5}
\end{figure}

\subsection{State preparations to the excited state}
\label{sec:state_preparation}

In this section, we will investigate the preparation of the system state to its
excited state with an external field using Initial-A and Initial-C
states as initial states. This allows us to see the effects of initial
correlations on the open quantum system dynamics during the state
preparation process. Let us first look at the case of the excited
state preparation in the closed quantum system (isolated from the
environment). The driving Hamiltonian for the state preparation pulse
is given in Eq.~(\ref{eq:Hd}) and a commonly used form with the RWA is 
\begin{equation}
H_{d}^{{\rm RWA}}(t)=\frac{\Omega_{R}}{2}\left(\sigma_{+}e^{-i\omega_{L}t}+\sigma_{-}e^{+i\omega_{L}t}\right).\label{eq:15}
\end{equation}
The RWA works well when the system is driven at resonance or near
resonance and the driving strength is weak, that is, when $\omega_{L}\cong2\Omega$
and $\Omega_{R}\ll\Omega$. 
We will determine the condition for the onset of RWA correction for
the excited state preparation later.

For the purpose of the excited state preparation by a
single sinusoidal field with frequency $\omega_L$ and amplitude
$\Omega_R$ as in Eq.~(\ref{eq:Hd}), not all strength of
the field amplitude 
$\Omega_R$ can 
prepare the qubit system from the ground state to the excited
state efficiently and effectively even in the ideal unitary case.   
When the relevant energy scales of the seemingly simple driven qubit
system are in the same order of magnitude, i.e., $\Omega_R\sim
\omega_L\sim 2\Omega$, the dynamics becomes complex. 
Figures \ref{fig:2}(a) and  \ref{fig:2}(b) show the time evolutions
of $\left\langle \sigma_{z}\left(t\right)\right\rangle $ of the qubit system
driven at the resonance frequency $\omega_{L}=2\Omega$ with amplitudes
$\Omega_R=2.4 \Omega$ and $\Omega_R=5.0\Omega$, respectively. The
complex dynamics in these cases are in contrast with the simple ideal unitary 
dynamics in the weak-amplitude rotating-wave cases as shown in the
inset of Fig.~\ref{fig:3}(a) for initial-D ground state. 
The first maximum in Fig.~\ref{fig:2}(a) is not the global maximum
thus require more time to complete the excited state preparation. This
is not efficient and is also more vulnerable when the environment
decoherence effect is taken into account.  
The first maximum in Fig.~\ref{fig:2}(b) is the global maximum but does
not get close to the desired accuracy of the ideal excited 
state of $\left\langle \sigma_{z}\right\rangle=1 $. 
Similar complex dynamics are also observed for $\Omega<\Omega_{R}<
10\Omega$ at resonance, $\omega_{L}=2\Omega$.
Thus in the following, we confine the amplitude $\Omega_R$ of the
field for the excited state
preparation to be in $0<\Omega_{R}\le \Omega$ or $\Omega_R>10 \Omega$.

To demonstrate the conditions for the breakdown of RWA, we again consider
the unitary case first. The RWA predicts a resonant and weak driving field
can transfer the system  from
the ground state to
the excited state in
the unitary case
with time duration $\Omega_{R}t=\pi$, the so-called $\pi$ pulse.
In the Bloch sphere representation the system state travels from the
south pole $\left|0\right\rangle \left\langle 0\right|$ ground state to
the north pole $\left|1\right\rangle \left\langle 1\right|$ excited state
with a time $t=\pi/\Omega_{R}$ \cite{key-01}.
The unitary time
evolutions of $\left\langle \sigma_{z}\left(t\right)\right\rangle $
and the corresponding Bloch sphere representation of the qubit system
for the state preparation from the ground state
to the excited state of the system
driven at resonance frequency $\omega_{L}=2\Omega$ with and without
the RWA denoted as Initial-DR (red cross) and Initial-D (black
circle), respectively, for different driving amplitudes 
are shown in Fig.~\ref{fig:3}. 
The deviation in the operation time between
the cases with and without the RWA increases as $\Omega_{R}$ increases.
As shown in Fig.~\ref{fig:3}(a) and (b), the
RWA operation time $t=\pi/\Omega_{R}$ to transfer the two-level system
from the ground state to the excited state is good for $\Omega_{R}\leq\Omega$,
but the state evolutions with and without the RWA for $\Omega_{R}=\Omega$
already show considerable difference. For larger value of $\Omega_{R}$,
the precise operation time that may be hard to calculate analytically
can be deduced numerically when the system state arrives at a place
nearest to the north pole. We note that a significant deviation of
the operation time from the RWA operation time of $t=\pi/\Omega_{R}$
can be observed for $\Omega_{R}\geq10\Omega$. One can see from
Fig.~\ref{fig:3}(c) 
that the actual operation time is about half of $\pi/\Omega_{R}$
for $\Omega_{R}=40\Omega$. This deviation clearly reflects the failure
of the RWA. 
Of course, this unitary
evolution is an ideal case. In reality, it is impossible to completely
isolate the system from the environment, so the travel time of $\pi$
pulse should be short enough to reduce the effect from the environment
such that the arrival position is close to the north pole \cite{key-25}.
Increasing the driving field amplitude $\Omega_{R}$ to shorten the
flight time is easier for experiments to perform than changing the
system-environment coupling to reduce the environment-induced decoherence
or/and decay. But the larger the $\Omega_{R}$ is, the less accurate
the RWA is.

To see the effect and the overall trend of the RWA and the initial
system-environment correlation on the
general problem of state preparation, we give in Fig.~\ref{fig:4}
 the errors of the prepared excited states as
functions of driving strength $\Omega_{R}$ and the system-bath
coupling constant $\xi$ for pulses without and with RWA and for
correlated and factorized initial states. 
The error of the prepared excited  state
is defined in Eq.~(\ref{eq:error_ES}).
Figure \ref{fig:4}(a) shows the error of the prepared states 
through the non-RWA pulse of the Hamiltonian of Eq.~(\ref{eq:Hd}) with
an initial state of the Initial-A state.  
The pulse duration is determined for the system state 
to reach the minimal error 
at a time around the RWA $\pi$-pulse time of $t=\pi/\Omega_{R}$
in the unitary evolution case.
One can see from Fig.~\ref{fig:4}(a) that a higher
prepared state fidelity or equivalently a
smaller error is achieved as $\xi$ becomes smaller and/or $\Omega_{R}$
becomes larger. However, for $\xi \leq 10^{-3}$,
 the trend of the error increases as $\Omega_{R}$ approaches $\Omega$.
This shows that $\Omega_{R}\rightarrow\Omega$ the somehow complicated
non-RWA dynamics raises the error. 
To compare with the state preparation by the RWA pulse, we gives in
Fig.~\ref{fig:4}(b) the error difference between
the error in Fig.~\ref{fig:4}(a) and the one by the RWA
 $\pi$ pulse of Eq.~(\ref{eq:15}), both starting from the Initial-A
state. 
If we require the error difference smaller than $10^{-2}$ as a
criterion for good agreement between the RWA and non-RWA, then the
onset of the non-RWA corrections
as seen from Fig. \ref{fig:4}(b) takes place at 
about $\Omega_{R}\sim 0.1\Omega$.   

We can calculate the pulse duration that gives the minimum error
between the prepared and target excited states in the open system
case. This fine-tuned pulse duration mimics the pulse length an
experimentalist wishes to achieve in a realistic situation. 
The resultant errors plotted in    
Fig.~\ref{fig:4}(c) with slightly smaller error values 
are quite similar to those of
Fig.~\ref{fig:4}(a) whose pulse durations are calculated in
the unitary closed system case. 
Fig.~\ref{fig:4}(d) gives the error difference between
the errors in Fig.~\ref{fig:4}(c) and 
the corresponding errors  with the same setup and pulse duration
as in Fig.~\ref{fig:4}(c) but with a different initial
state, the factorized Initial-C state. 
Since the pulse duration is adopted for the Initial-A to reach the
minimum error, 
the error
difference in Fig.~\ref{fig:4}(d) is mostly negative.
One can also observe that the error difference between the cases of 
Initial-A and Initial-C start to
emerge at about $\xi\geq 10^{-2}$. This is due to the fact that 
the initial system-bath correlation is
proportional to $\xi$, so when $\xi$ reach a value of about $10^{-2}$,
the error difference between correlated and factorized initial states 
becomes visible.  
Furthermore, the error difference increases as $\Omega_{R}$ increases.
This is because small $\Omega_{R}$ requires longer preparation time
so that the Initial-C state has enough time to establish
system-bath correlations and then both of the initial states approach
the same prepared states. However, as  
$\Omega_{R}$ increases, the shorter preparation time with smaller
pulse duration makes the difference between the corresponding prepared
states of the two initial states become more appreciable.

We choose
a larger coupling constant of $\xi=0.1$ 
for the time evolution and the
corresponding Bloch sphere plots in 
Fig.~\ref{fig:3}
in order to show the breakdowns of the factorization approximation for the
initial system-environment state in the open quantum system. 
The operation times to prepare the excited state are chosen to be
the same as the ideal unitary cases in order for comparison. One can see from
Fig.~\ref{fig:3}(b) that after applying the
state preparation pulses, the system states, except those in the unitary
case, are all far off the target excited state $\left|1\right\rangle \left\langle 1\right|$,
not to mention those of a smaller pulse amplitude in Fig.~\ref{fig:3}(a).
Furthermore, the difference in $\left\langle \sigma_{z}(t)\right\rangle $
between evolutions with and without the RWA for every initial state
in Fig.~\ref{fig:3}(c) is clearly visible.
Observing from Figs.~\ref{fig:3}(a) and (d)
as well as from Figs.~\ref{fig:3}(b) and (e),
one can conclude that the pulse amplitudes of $\Omega_{R}=0.2\Omega$
and $\Omega_{R}=\Omega$ are too small to transfer those initial states
to the excited state for the open quantum system with coupling constant
$\xi=0.1$. Thus larger pulse amplitudes are required as in Fig.~\ref{fig:3}(c)
with a large value of $\Omega_{R}=40\Omega$.
Figure \ref{fig:3}(f)
is the evolution trajectory in the Bloch sphere representation for
the state preparation process at $\Omega_{R}=40\Omega$ of Fig.~\ref{fig:3}(c).
The resultant prepared states 
with the operation time about $0.5\pi/\Omega_{R}$
are much closer to the excited state at the north pole of the Bloch
sphere than those of the open system cases in 
Figs.~\ref{fig:3}(d)and (e).

Recognizing already the failure of the RWA Hamiltonian at large pulse
amplitudes, we consider only the driving field Hamiltonian without
the RWA and plot in Fig.~\ref{fig:5} the
dynamics of 
the trace distance between the states evolving from the 
Initial-A and Initial-C states (abbreviated as ``A \& C'')
for the state preparation process
with the pulse duration time determined by minimizing the 
error between the target excited state and the prepared excited state
evolving from the Initial-A state for each value of $\Omega_{R}$. 
It is clear that as the pulse amplitudes  $\Omega_{R}$ increases,
the operation times for the excited state preparation in the open
system become shorter and the resultant prepared states are closer
to the excited state $\left\langle \sigma_{z}\right\rangle =1$. The
differences between $\left\langle \sigma_{z}(t)\right\rangle $ evolving
from the correlated Initial-A state and from the factorized Initial-C state
in Figs.~\ref{fig:3}(a), (b), and (c) are hard
to recognize, 
but the
corresponding difference in terms of the trace distance can be easily
observed in Figs.~\ref{fig:5}(a), (b)
and (c), respectively, indicating that the trace distance is indeed
a sensitive and appropriate measure. We note here that the trace distances
 under a driving
field in Figs.~\ref{fig:5}(a) and (b)
increase above their initial values, but the trace distance 
for ``A \& C'' in
Figs.~\ref{fig:5}(c) (red solid line)
decreases initially below its initial value. 

The trace distances shown in Fig.~\ref{fig:5}
also illustrate the difference between the system states evolving
from either the Initial-A state (black dashed line) or 
the Initial-C state (blue dotted line) and from the
ideal unitary evolution of the Initial-D state. One can clearly see
that for $\Omega_{R}=40\Omega$, the values of the trace distance  
between states evolving from the Initial-A and from the Initial-D
state (black dashed line) in the whole operation time in
Fig.~\ref{fig:5}(c) are around 
a small value of order of $10^{-3}$ and they are slightly higher for the
cases with the Initial-C state at the final time (blue dotted line). 
These small values
of the trace distances in Fig.~\ref{fig:5}(c)
as compared to those in the inset of Fig.~\ref{fig:5}(a)
and in Fig.~\ref{fig:5}(b) demonstrate
that large pulse amplitudes are necessary to perform the excited state
preparation with the correlated initial state for the system-environment
coupling parameter $\xi=0.1$.

\begin{figure*}
\includegraphics[width=1\textwidth]{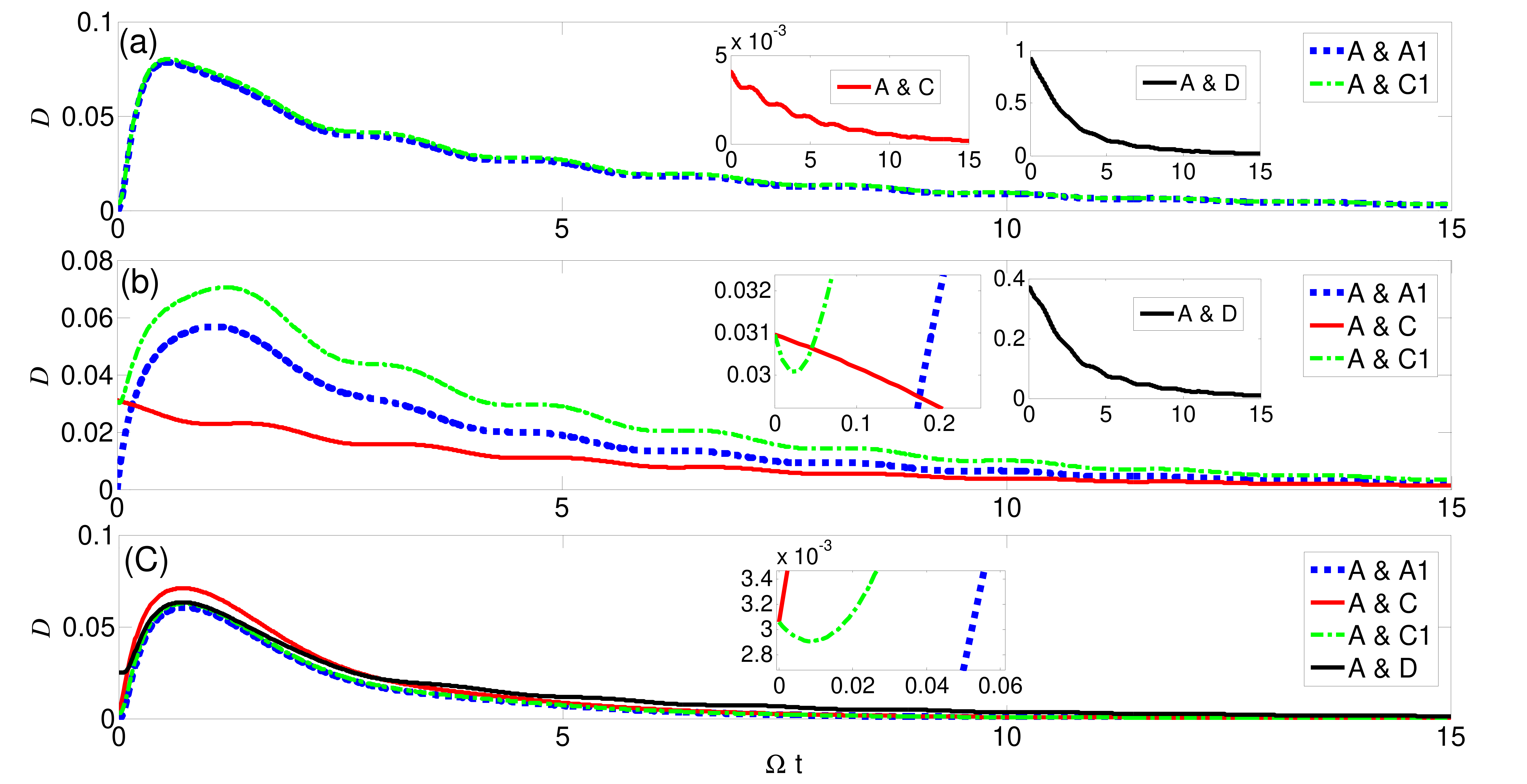}
\caption{(Color online) Field-free evolutions of the trace distances between the reduced system
states evolving from several kinds of prepared states for different
strengths of (a) $\Omega_{R}=0.2\Omega$, (b) $\Omega_{R}=\Omega$,
(c) $\Omega_{R}=40\Omega$.
The zoom-in plots of the initial evolutions of (b) and (c) are shown 
in their respective insets.
The time evolution of the trace distance between the reduced system states
evolving from the Prepared-A and Prepared-A1 states without the external
field is abbreviated as ``A \& A1'' (blue dotted line). 
Similar abbreviations for other time evolutions of the
trace distance evolving from their respective prepared
states are applied accordingly. 
Other parameters are $\xi=0.1$, $\omega_{c}=7.5\Omega$ and
$k_{B}T=1/\beta=0.1\Omega$.}
\label{fig:6} 
\end{figure*}

\begin{figure*}
\includegraphics[width=1\textwidth]{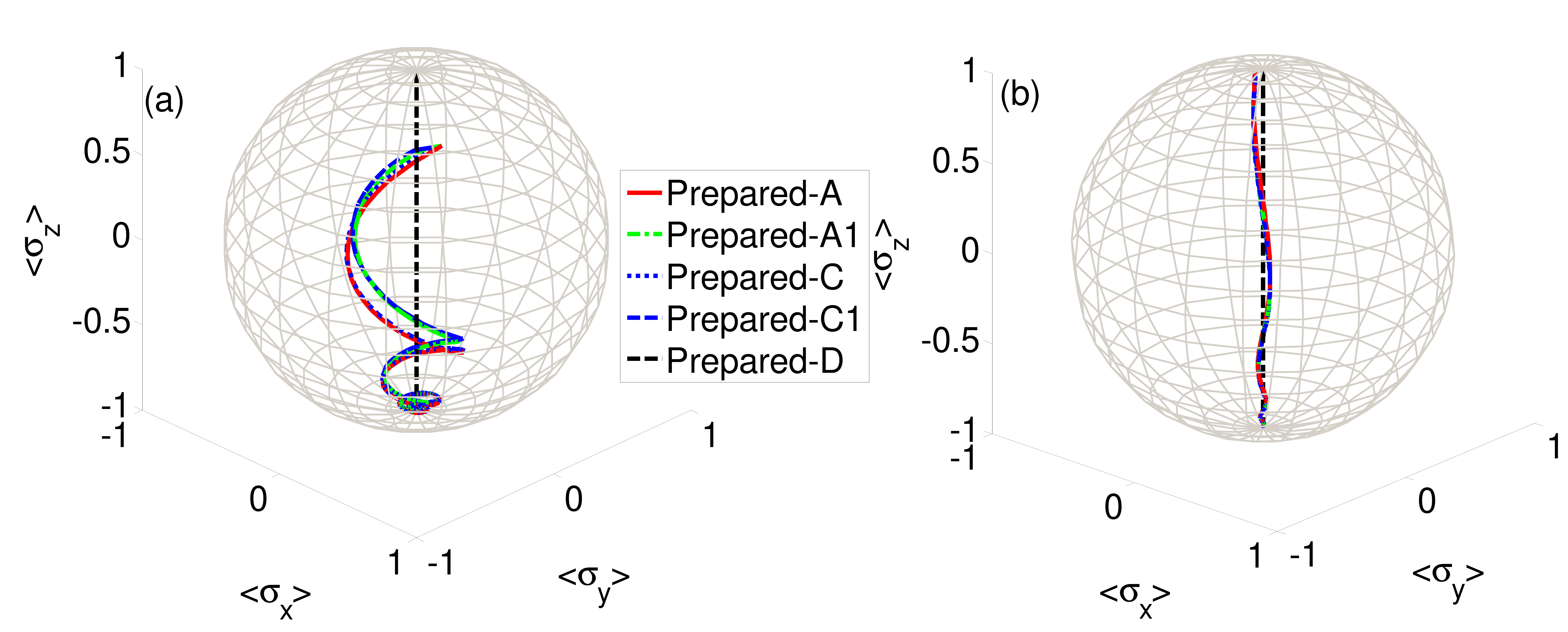}
\caption{(Color online) Field-free evolutions of the trajectories evolving from the north
to the south direction in the Bloch sphere representation with different
prepared states as initial states for different pulse strengths of
(a) $\Omega_{R}=\Omega$, and (b) $\Omega_{R}=40\Omega$, corresponding to
the time evolutions of the system in Fig.~\ref{fig:6}(b) and
(c), respectively. }
\label{fig:7} 
\end{figure*}

\subsection{Non-factorized prepared states after the system state preparation}
\label{sec:prepared_states}

We next consider the dynamics of the trace distance between these
correlated states and their corresponding factorized states in a subsequent
field-free evolution. In other words, we take the resultant correlated
system-environment states after the state preparation as the initial
states for the subsequent field-free evolution. To this end, the values
$\rho_{s}(t)$ and $\mathcal{K}_{k}(t)$ evolving from the Initial-A
and Initial-C states through Eqs. (\ref{eq:11}) and (\ref{eq:12})
at the final time of Figs.~\ref{fig:5}(a),
(b) and (c) (i.e., at the times when the external pulses are turned
off) are obtained and denoted as corresponding Prepared-A and Prepared-C
states, respectively. Unless there exists possible confusion, we will,
for convenience, simply use letters A and C in the following state
superscripts and figure legends to represent the Prepare-A and Prepared-C
states, respectively. For example, the Prepared-A state is represented
by $\rho_{s}^{A}=tr_{b}\left(\rho_{T}^{A}\right)$ and $\mathcal{K}_{k}^{A}$,
and similar notations are applied to the Prepared-C state and other
states introduced later. Specifically, the initial input states we
will consider in the decomposition by $P$ and $Q$ projection operators
for the subsequent field-free evolution are (i) Prepared-A state:
$\rho_{T}^{A}$, correlated state with the reduced system state in
the factorized part of the state decomposition being $\rho_{s}^{A}=tr_{b}\left(\rho_{T}^{A}\right)$
and the non-factorized part being $\mathcal{K}_{k}^{A}$. (ii) Prepared-A1
state: the factorized part of the Prepared-A state, $\rho_{T}^{A1}=\rho_{s}^{A}\otimes\rho_{b}$,
implying $\mathcal{K}_{k}^{A1}=0$. (iii) Prepared-C state: $\rho_{T}^{C}$,
a correlated state with the reduced system state in the factorized
part being $\rho_{s}^{C}=tr_{b}\left(\rho_{T}^{C}\right)$ and the
non-factorized part being $\mathcal{K}_{k}^{C}$. (iv) Prepared-C1
state: the factorized part of the Prepared-C state, $\rho_{T}^{C1}=\rho_{s}^{C}\otimes\rho_{b}$,
implying $\mathcal{K}_{k}^{C1}=0$. (v) Prepared-D state, the ideal
prepared factorized state $\left|1\right\rangle \left\langle 1\right|\otimes\rho_{b}$,
implying $\mathcal{K}_{k}^{D}=0$.

The results of the trace distance between the reduced system states
evolving from these correlated states and their corresponding factorized
states in the field-free evolution are presented in Fig.~\ref{fig:6}.
The Prepared-A, -C and -D states in Fig.~\ref{fig:6}
are abbreviated simply as A, C and D states, respectively. The other
factorized states of the prepared states are similarly abbreviated.
One can find that the values of the trace distance, e.g., ``A \&
C'' at the initial time in Figs.~\ref{fig:6}(a)-(c), 
have the same values as those at the final time in Figs.~\ref{fig:5}(a)-(c),
respectively. The trace distance for ``A \& C'' (red solid line) shown
in the inset of Fig.~\ref{fig:6}(a) is small,
less than $5\times10^{-3}$, indicating that under the external field
with a small pulse amplitude of $\Omega_{R}=0.2\Omega$, the long
pulse duration allows the factorized Initial-C state to establish
system-bath correlations such that Prepared-A state and Prepared-C
state are almost the same. The decrease of the trace distance from
the tiny initial value of the field-free evolution to zero shows that
the Prepared-A and Prepared-C states are approaching the same equilibrium
state. The trace distance for ``A \& C'' (red solid line) during the
course of the 
excited state preparation with $\Omega_{R}=40\Omega$ is also small
as shown in Fig.~\ref{fig:5}(c), but
this trace distance grows above its initial value in Fig.~\ref{fig:6}(c)
to larger values in the field-free evolution in contrast to the decrease
in the inset of Fig.~\ref{fig:6}(a).
This is because the short duration of the strong preparation pulse
of $\Omega_{R}=40\Omega$ makes the factorized Initial-C state evolve
only a little bit away from its original state, i.e., just in the
beginning stage to establish some system-environment correlation.
As a result, one can observe in Fig.~\ref{fig:6}(c) 
that the trace distance for ``A \& C'' (red solid line) increases
initially and then 
to a peak value. After that, it decreases and finally approaches zero
at long times when the system-bath correlations of ``A \& C'' both
reach equilibrium.

The trace distance between the reduced system states evolving from
the Prepared-A and its corresponding factorized part of Prepared-A1
state (blue dotted line) increases above its initial value for any
and all parameter 
settings in Fig.~\ref{fig:6}. Because
the Prepared-A and Prepared-A1 states are in the decomposition of
the same initial bath reference state $\rho_{b}$, the increase of
the trace distance between them indicates clearly the existence of
the initial system-environment correlation. However, the time evolutions
of the trace distances other than those of ``A \& A1'' do not always increase
above their initial values. See, for example, the zoom-in plots of the
initial time evolutions the trace distance ``A \& C1'' in green
dot-dashed line
in the insets of 
Figs.~\ref{fig:6}(b) and (c). 
We will discuss in some more details 
the various dynamical
behaviors after we introduce bounds for the trace distance in  
Sec.~\ref{sec:bounds}.
The evolution trajectories from the north to the south direction in the Bloch
sphere representation with different prepared states as initial states
for pulse strengths of $\Omega_{R}=\Omega$ and $\Omega_{R}=40\Omega$
corresponding to those in Fig.~\ref{fig:6}(b) and
(c), respectively, are shown in Fig.~\ref{fig:7}.

\begin{figure*}
\includegraphics[width=1\textwidth]{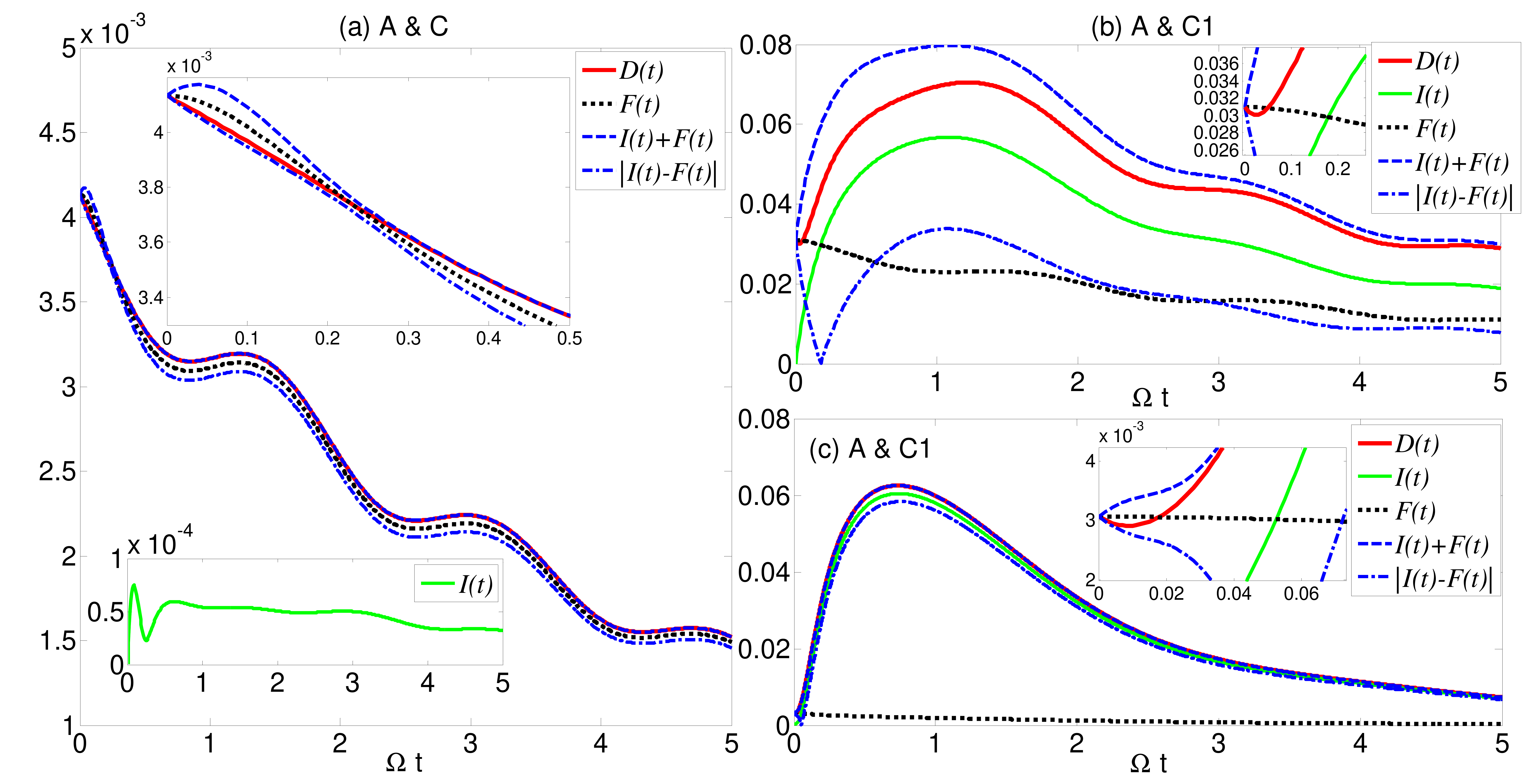}
\caption{(Color online) Field-free evolutions of the trace distance (red solid line) and corresponding
upper (blue dashed line) and lower (blue dot-dashed line) bounds in
Eq.~(\ref{eq:17}).
The trace distances ${D}(t'=0,t,\rho^{A,C\left(C1\right)})$  
between the reduced system states evolving from the Prepared-A and
Prepared-C(Prepared-C1) states after the application of the preparation
pulse are abbreviated
as ${D}(t)$  for ``A \& C(C1)'' in red solid lines.
The different pulse amplitudes and trace distances shown are 
(a) ``A \& C'' with $\Omega_{R}=0.2\Omega$, 
(b)  ``A \& C1'' with $\Omega_{R}=\Omega$ and (c)
 ``A \& C1'' with $\Omega_{R}=40\Omega$. 
The zoom-in plots of the initial evolutions of (a), (b) and (c) are shown 
in their respective insets. 
Here the time $t'$ when the external preparation pulse is
turned off is relabeled as $t'=0$. The time evolutions of the quantities
$I(0,t,\rho^{A,C\left(C1\right)})$ and $F(0,t,\rho^{A,C\left(C1\right)})$
for ``A \& C(C1)'' abbreviated as $I(t)$ (green solid line) and
$F(t)$ (black dotted line), respectively, are also presented. }
\label{fig:8} 
\end{figure*}

\section{Bounds for the trace distance: role of the system-environment correlations }
\label{sec:bounds}
To gain a quantitative understanding of the diverse
behavior of the trace distance in general, we present an analysis
that bounds the finite-time difference in trace distance by sharply
defined quantities that link to the existence of the system-environment
correlation in Sec.~\ref{sec:ULbounnds}. We will then take the
field-free case in Fig.~\ref{fig:6}
as an example to find and analyze the upper and lower bounds
in Sec.~\ref{sec:bound_example}. Similar
bounds using another form of the initial bath reference state are
discussed in Ref.~\cite{key-30}.

\subsection{Upper and lower bounds}
\label{sec:ULbounnds}

The trace distance between two different reduced system states $\rho_{s}^{\alpha}\left(t',t\right)$
and $\rho_{s}^{\beta}\left(t',t\right)$ of a given open quantum system
evolving from time $t'$ to $t$ can be written as 
\begin{eqnarray}
{D}(t',t,\rho^{\alpha,\beta}) & = & {D}\left(\rho_{s}^{\alpha}\left(t',t\right),\rho_{s}^{\beta}\left(t',t\right)\right)\nonumber \\
 & = & \frac{1}{2}\left\Vert tr_{b}\left\{ \mathcal{G}_{T}\left(t,t'\right)\Delta\rho_{T}^{\alpha,\beta}\left(t'\right)\right\} \right\Vert ,\label{eq:DT}
\end{eqnarray}
where $\Delta\rho_{T}^{\alpha,\beta}\left(t'\right)=\rho_{T}^{\alpha}\left(t'\right)-\rho_{T}^{\beta}\left(t'\right)$.
Any joint system-environment state $\rho_{T}(t)$ can be decomposed
by the $P$ and $Q$ projection operators as the expression of Eq.~(\ref{eq:rhoT}).
Thus one has $\Delta\rho_{T}^{\alpha,\beta}\left(t'\right)=\Delta\rho_{s}^{\alpha,\beta}\left(t'\right)\otimes\rho_{b}+Q\Delta\rho_{T}^{\alpha,\beta}\left(t'\right)$,
where $\Delta\rho_{s}^{\alpha,\beta}\left(t'\right)=\rho_{s}^{\alpha}(t')-\rho_{s}^{\beta}(t')=tr_{b}[\rho_{T}^{\alpha}\left(t'\right)-\rho_{T}^{\beta}\left(t'\right)]$.
Substituting the above expression into Eq.~(\ref{eq:DT}) and using
the triangular inequality for the trace norm 
\begin{equation}
\left\Vert \Gamma\right\Vert +\left\Vert \Gamma'\right\Vert \geq\left\Vert \Gamma+\Gamma'\right\Vert \geq\left|\left\Vert \Gamma\right\Vert -\left\Vert \Gamma'\right\Vert \right|,\label{eq:Trianngular}
\end{equation}
where $\Gamma$ and $\Gamma'$ are linear trace class operators, we
obtain bounds to the trace distance ${D}(t',t,\rho^{\alpha,\beta})$:
\begin{eqnarray}
 &  & I(t',t,\rho^{\alpha,\beta})+F(t',t,\rho^{\alpha,\beta})\nonumber \\
 & \geq & {D}(t',t,\rho^{\alpha,\beta})\nonumber \\
 & \geq & \left|I(t',t,\rho^{\alpha,\beta})-F(t',t,\rho^{\alpha,\beta})\right|,\label{eq:17}
\end{eqnarray}
where 
\begin{eqnarray}
F(t',t,\rho^{\alpha,\beta}) & = & \frac{1}{2}\left\Vert tr_{b}\left\{ \mathcal{G}_{T}\left(t,t'\right)\left[\Delta\rho_{s}^{\alpha,\beta}\left(t'\right)\otimes\rho_{b}\right]\right\} \right\Vert ,\label{eq:18}\\
I(t',t,\rho^{\alpha,\beta}) & = & \frac{1}{2}\left\Vert tr_{b}\left\{ \mathcal{G}_{T}\left(t,t'\right)\left[Q\Delta\rho_{T}^{\alpha,\beta}\left(t'\right)\right]\right\} \right\Vert .\label{eq:19}
\end{eqnarray}
The quantity $F(t',t,\rho^{\alpha,\beta})$ in Eq.~(\ref{eq:18})
is the trace distance between two reduced system states at time $t$,
which evolved from their respective total system-bath product states
at an earlier time $t'$. One can describe the time evolution of the
reduced system in $F(t',t,\rho^{\alpha,\beta})$ through a family
of completely positive dynamical maps. Completely positive maps with
the same initial environment state are contractive and they form a
time-dependent semigroup, so are Markovian quantum stochastic processes
\cite{Gorini76,Lindblad76}. This is why the concepts of no initial
correlation, complete positive maps, contractivity and Markovianity
are often linked or discussed all together in the literature. Contractivity
means that the distinguishability of two input states cannot increase
in time, i.e., the trace distance of the system density matrices of
the quantum process decreases with time. Thus one has $F(t',t,\rho^{\alpha,\beta})\leq F(t',t',\rho^{\alpha,\beta})$
due to the contractive property of the completely positive maps with
the same initial environment states. The quantity $I(t',t,\rho^{\alpha,\beta})$
in Eq.~(\ref{eq:19}) keeps track of the effect of the system-bath
correlation at a time $t'$ on the subsequent dynamics of the reduced
system at time $t$. Since the propagator superoperator $\mathcal{G}_{T}\left(t',t'\right)=1$
and the property $tr_{b}[Q\rho_{T}\left(t'\right)]=0$ (due to $PQ=0$),
one has at the initial propagating time $t'$ the quantity $I(t',t',\rho^{\alpha,\beta})=0$.
In general, $I(t',t,\rho^{\alpha,\beta})$ depends on the system-bath
correlations established at time $t'$ for each of $Q\rho_{T}^{\alpha}\left(t'\right)$
and $Q\rho_{T}^{\beta}\left(t'\right)$. If there are no system-bath
correlations at time $t'$, i.e., $Q\rho_{T}^{\alpha}\left(t'\right)=0=Q\rho_{T}^{\beta}\left(t'\right)$,
or if there are system-bath correlations but their difference vanishes
at time $t'$, i.e., $Q\Delta\rho_{T}^{\alpha,\beta}\left(t'\right)=0$,
then $I(t',t,\rho^{\alpha,\beta})=I(t',t',\rho^{\alpha,\beta})=0$.
Thus $I(t',t,\rho^{\alpha,\beta})>I(t',t',\rho^{\alpha,\beta})=0$
can serve as a witness of whether there is difference of the initial
correlations at time $t'$ between the two total states, i.e., whether
$Q\Delta\rho_{T}^{\alpha,\beta}\left(t'\right)\neq0$.

Witnesses for non-Markovianity through detecting deviations from contractive
dynamics have been proposed and investigated \cite{key-15,key-24,key-28,key-26,Dajka11}.
Based on these investigations, if a process is not contractive (or
the trace distance of a process increases) at some instant time or
in some time intervals for some states, then the process is non-Markovian.
Thus it is instructive to find the necessary condition and the sufficient
condition for the increase of the trace distance in the time interval
$[t',t]$. Again, since $I(t',t',\rho^{\alpha,\beta})=0$, one has
${D}(t',t',\rho^{\alpha,\beta})=F(t',t',\rho^{\alpha,\beta})$. Thus
by using these relations, the necessary condition for the increase
of the trace distance ${D}(t',t,\rho^{\alpha,\beta})>{D}(t',t',\rho^{\alpha,\beta})$
is then the increase of the upper bound in Eq.~(\ref{eq:17}): 
\begin{equation}
I(t',t,\rho^{\alpha,\beta})+F(t',t,\rho^{\alpha,\beta})>I(t',t',\rho^{\alpha,\beta})+F(t',t',\rho^{\alpha,\beta}),\label{eq:20}
\end{equation}
and the sufficient condition is the increase of the lower bound in
Eq.~(\ref{eq:17}): 
\begin{eqnarray}
 &  & \left|I(t',t,\rho^{\alpha,\beta})-F(t',t,\rho^{\alpha,\beta})\right|\nonumber \\
 &  & >\left|I(t',t',\rho^{\alpha,\beta})-F(t',t',\rho^{\alpha,\beta})\right|.\label{eq:21}
\end{eqnarray}
If the value of $I(t',t,\rho^{\alpha,\beta})$ makes the sufficient
condition, Eq.~(\ref{eq:21}), satisfied for {\em some} choice
of $t'$, $t$, and $\rho^{\alpha}(t')$, $\rho^{\beta}(t')$, then
the dynamics is non-Markovian. If, however, the value of $I(t',t,\rho^{\alpha,\beta})$
makes the following condition, opposite to the necessary condition
of Eq.~(\ref{eq:20}), 
\begin{equation}
I(t',t,\rho^{\alpha,\beta})+F(t',t,\rho^{\alpha,\beta})<I(t',t',\rho^{\alpha,\beta})+F(t',t',\rho^{\alpha,\beta})\label{eq:21M}
\end{equation}
satisfied for {\em all} the $t'$, $t$, and $\rho^{\alpha}(t')$,
$\rho^{\beta}(t')$, the dynamics is Markovian.

Similar bounds of the finite-time difference in trace distance and
forms of the necessary condition and sufficient condition for the
increase of the trace distance can be found in Ref.~\cite{key-30}.
However, the decomposition of a general joint system-environment state
in Ref.~\cite{key-30} is through the reduced states of the system
and the reduced state of the environment \cite{Uchiyama12,key-30,key-31}:
\begin{equation}
\rho_{T}(t)=\rho_{s}\left(t\right)\otimes\rho_{e}(t)+\chi_{se}\left(t\right),\label{eq:rhoT_E}
\end{equation}
where $\rho_{e}(t)=tr_{s}[\rho_{T}(t)]$ is the reduced state of the
environment at time $t$ and $\chi_{se}(t)$ accounts for the remaining
correlations between the open system and the environment. This decomposition
is different from the decomposition of Eq.~(\ref{eq:rhoT}) using
the projection operators $P$ and $Q$. The reference bath state is
the fixed bath thermal equilibrium state $\rho_{b}$ for all total
system-environment states in our formulation, while the reduced environment
states $\rho_{e}(t)$ in Ref.~\cite{key-30} are time-dependent and
are different for different total system-environment states. As a
consequence, the remaining system-bath correlations in these two decompositions
are also different. The disadvantage of the decomposition of Eq.~(\ref{eq:rhoT_E})
is that it is hard to calculate the corresponding bounds presented
in Ref.~\cite{key-30} unless an exact dynamical solution of the
total system-bath state is available to evaluate the time-dependent
reduced state of the environment. In contrast, our decomposition of
Eq.~(\ref{eq:rhoT}) can be directly combined with the perturbative
master equation approach by the projection operator technique even
with strong external driving fields for the calculation of the dynamics
of the derived corresponding bounds of the trace distance.

\subsection{Examples of the field-free evolutions for the upper and lower bounds}
\label{sec:bound_example}

Next we present the time evolutions of the $I(t',t,\rho^{\alpha,\beta})$
and $F(t',t,\rho^{\alpha,\beta})$ of Eqs.~(\ref{eq:18}) and (\ref{eq:19})
for the upper and lower bounds in Eq.~(\ref{eq:17}). We will take
the diverse behaviors of the dynamics of the trace distances presented
in Fig.~\ref{fig:6} as examples to
illustrate that they are indeed within the bounds. Compared to Eqs.~(\ref{eq:4})
and (\ref{eq:DT}), Eqs.~(\ref{eq:18}) and (\ref{eq:19}) satisfy
the same equation of motion of Eq.~(\ref{eq:5}) as Eq.~(\ref{eq:DT})
but with different initial conditions. To second order in system-environment
interaction Hamiltonian, this same equation of motion can be cast
into the master equation of Eqs.~(\ref{eq:11}) and (\ref{eq:12})
in the extended auxiliary Liouville space. The initial condition for
$F(t',t,\rho^{\alpha,\beta})$ is $F(t',t',\rho^{\alpha,\beta})=\frac{1}{2}\left\Vert \Delta\rho_{s}^{\alpha,\beta}\left(t'\right)\right\Vert $
and the initial inputs in terms of the extended Liouville space formulation
become $\rho_{s}^{\alpha,\beta}\left(t'\right)$ and $\mathcal{K}_{k}^{\alpha,\beta}\left(t'\right)=0$.
Similarly, the initial condition for $I(t',t,\rho^{\alpha,\beta})$
is $I(t',t',\rho^{\alpha,\beta})=0$ and in the extended Liouville
space formulation, the initial inputs are $\rho_{s}^{\alpha,\beta}\left(t'\right)=0$
and $\mathcal{K}_{k}^{\alpha,\beta}\left(t'\right)$ that correspond
to $Q\rho_{T}^{\alpha,\beta}(t')$.

Figure \ref{fig:8} shows the dynamics of the
trace distance ${D}(t'=0,t,\rho^{A,C(C1)})$ between the reduced system
states evolving from the Prepared-A and Prepared-C(Prepared-C1) states
in the field-free case after the application of the preparation pulse
with different pulse amplitudes. We have set the time when the external
preparation pulse is turned off as $t'$ and relabeled it as $t'=0$.
We also plot in Fig.~\ref{fig:8} the quantities
$I(0,t,\rho^{A,C(C1)})$ (green solid line), $F(0,t,\rho^{A,C(C1)})$
(black dotted line) and the corresponding
upper (blue dashed line) and lower (blue dot-dashed line) bounds of
the trace distance.  
The dynamical behaviors of the trace distances shown in 
Fig.~\ref{fig:6} 
and Fig.~\ref{fig:8} are quite diverse.
The time evolution of
the trace distance of ``A \& C'' in red solid line in 
Fig.~\ref{fig:8}(a)
shows non-monotonic decrease from their initial values
(indicating a witnesses of non-Markovianity), 
while that in Fig.~\ref{fig:6}(c) increases above its initial value
(indicating a witness of initial correlation), reaches a
peak and then decreases. 
The trace distances of ``A \& C1'' in red solid lines  
in Figs.~\ref{fig:8}(b)
and (c) decrease at first below their initial values  
(see the zoom-in plots of the
initial time evolutions in the insets), but
then increase above their initial values and later decrease again.
One can observe that the time evolutions of the trace distance, no
matter how diverse their behaviors are, are all between the bounds
satisfying Eq.~(\ref{eq:17}). 
For example, the trace distance ${D}(0,t,\rho^{A,C1})$ in 
Fig.~\ref{fig:8}(c) 
is close to the upper bound for almost all the period of time shown but is still within
the bounds. 
The question about whether or how
the trace distances are close to the upper or lower bounds depends on
what the two 
reduced system states and the non-factorized parts are in Eq.~(\ref{eq:17}).
We find that the trace
distance ${D}(0,t,\rho^{A,C1})$ in Fig.~\ref{fig:8}(b)
will change its behavior from being around the middle of the bounds
to oscillating closely to the upper or the lower bound if one changes
the duration time of the preparation pulse prior to the field-free
evolution with other conditions and parameters unchanged.

Even though the effect of the system-environment correlation increases
$I(0,t,\rho^{\alpha,\beta})>I(0,0,\rho^{\alpha,\beta})=0$, the trace
distance between the reduced system states does not always increase
above its initial value. An increase of the trace distance
(distinguishability) between
the reduced states, however, requires $I(0,t,\rho^{\alpha,\beta})$
to prevail over the difference of $[F(0,0,\rho^{\alpha,\beta})-F(0,t,\rho^{\alpha,\beta})]$,
the amount of contraction due to the completely positive maps. Indeed,
whenever ${D}(0,t,\rho^{\alpha,\beta})$ increases above its initial
values in Figs.~\ref{fig:8}(b) and (c), the
necessary condition, Eq.~(\ref{eq:20}), is always satisfied. It
is only when the initial $F(0,0,\rho^{\alpha,\beta})=0$ (implying
$F(0,t,\rho^{\alpha,\beta})=0$) that the trace distance increases
above its initial value once $I(0,t,\rho^{\alpha,\beta})>0$. This
follows from the sufficient condition, Eq.~(\ref{eq:21}). If the
two initial states $\rho^{\alpha}$ and $\rho^{\beta}$ are chosen
to be, respectively, a correlated state and its corresponding factorized
state with the same reduced system state such as the Prepared-A and
Prepared-A1 states, then $F(0,0,\rho^{\alpha,\beta})=0=F(0,t,\rho^{\alpha,\beta})$
and $I(0,t,\rho^{\alpha,\beta})>0$ {[}note that
$I(0,t\rightarrow\infty,\rho^{\alpha,\beta})\to 0${]}. 
This explains why in Fig.~\ref{fig:6},
an increase of the trace distance ${D}(0,t,\rho^{A,A1})$ over its
initial value always takes place. There exist some certain times (except
the times $\Omega t<0.02$) in Fig,~\ref{fig:8}(c)
that $I(0,t,\rho^{A,C})$ not only greater than zero but also large
enough to make the sufficient condition, Eq.~(\ref{eq:21}), satisfied.
Thus in this case, ${D}(0,t,\rho^{A,C})$ increases over its initial
value.

\section{CONCLUSION}

\label{sec:Conclusion}

We have investigated the problem of system state preparation by an
external field in the presence of an environment with initial system-environment
correlations. The open quantum system model we consider is a spin-boson
model and the target system state we wish to prepare is the excited
state of the two-level system (or the spin). Starting with an initial
joint system-environment state in the correlated total thermal equilibrium
state, we use the projection operator technique to obtain a perturbative
time-nonlocal master equation for the reduced system density matrix,
which takes into account the effect of the initial system-environment
correlation. To describe the dynamics of the system under the application
of a time-dependent external (strong) field, one would need to solve
the master equation that is a time-nonlocal, time-ordered
integro-differential
equation. Instead of solving the quantum master equation directly,
we express the bath correlation function in a multi-exponential form
and transform the time-nonlocal and time-ordered integro-differential
equation into a set of coupled linear time-local differential equations
in an extended auxiliary Liouville space. The resultant coupled differential
equations that are time-local and have no time-ordering and memory
kernel integration problems are much easier to solve. Moreover, these
coupled equations take into account the initial system-environment
correlation and can deal with the case of strong driving fields with
which the RWA breaks down.

We have used the time evolutions of $\langle\sigma_{z}(t)\rangle$,
the trace distance and the trajectory in the Bloch sphere representation
to study the effects of the initial system-environment correlations, the
amplitudes of the preparation pulses and the strengths of system-environment
couplings on the system state preparation. 
We have found that unless
the system-bath coupling is very weak as compared to the pulse amplitude
$\Omega_{R}$, it is hard to perform the high-fidelity system state
preparation with a single resonant sinusoidal field.
For the model and parameters discussed in this paper, when the
system-bath coupling constant $10^{-5}\xi \leq 10^{-3}$ and 
$0.01<(\Omega_R/\Omega)<0.8$, the error of the excited state
preparation can be less than $10^{-2}$. 
For a slightly larger value of
the system-bath coupling and considering the initial system-environment
state to be in the total thermal equilibrium state, a larger value
of $\Omega_{R}$ is required. When $\Omega_{R}$ is not very small
comparing to the frequency of the two-level system $\Omega$, the
commonly adopted RWA fails. We have found that the onset of the
non-RWA corrections takes
place at about $\Omega_R\sim 0.1\Omega$ where the error difference of the
excited state
preparation between the RWA and non-RWA cases is about $10^{-2}$.
We have also investigated the case when $\Omega_{R}$ is much larger
than $\Omega$. 
We have also found that the state preparation error of 
the initial factorization approximation for the total
system-bath states starts to show deviation from that of 
the correlated case when the system-bath coupling
constant $\xi>10^{-2}$.

We have introduced a more computable upper and lower bounds for the
trace distance between two reduced system states in a decomposition
of the total system-environment states with a fixed reference thermal
environment state by the projection operators $P$ and $Q$. These
bounds give a sufficient condition and a necessary condition for the
increase of the trace distance and are related to the witnesses for
non-Markovianity and for the difference in initial system-environment
correlations. We have used these upper and lower bounds to describe
the diverse behaviors of the field-free time evolutions of the trace
distance between reduced system states evolving from various correlated
and uncorrelated states right after the state preparation pulses are
turned off. These upper and lower bounds that can be calculated
through the perturbative master equation approach can be directly applied to
a wide range of open system models to study problems, such as the state
distinguishability, the   
effects of initial system-bath correlations and their corresponding dynamics.

\begin{acknowledgments}
We acknowledge support from the the Ministry of Science and Technology
of Taiwan under Grant No.~103-2112-M-002-003-MY3, from the National
Taiwan University under Grants No.~NTU-ERP-104R891402, and from the
thematic group program of the National Center for Theoretical Sciences,
Taiwan.\end{acknowledgments}


\begin{thebibliography}{10}
\bibitem{key-01}H. J. Carmichael, \textit{Statistical Methods in
Quantum Optics 1} (Springer, Berlin, 1999).

\bibitem{Breuer02} H. P.~Breuer and F.~Petruccione, \textit{The
Theory of Open Quantum Systems} (Oxford University Press, Oxford,
2002). 

\bibitem{Rivas12}\'{A}. Rivas and S. F. Huelga, \textit{Open Quantum
Systems: An Introduction} (Springer, Heidelberg, 2012).

\bibitem{key-E1}C. F. Li, J. S. Tang, Y. L. Li, and G. C. Guo, Phys.
Rev. A \textbf{83}, 064102 (2011).

\bibitem{key-E2}A. Smirne, D. Brivio, S. Cialdi, B. Vacchini, and
M. G. A. Paris, Phys. Rev. A \textbf{84}, 032112 (2011).

\bibitem{key-E3}M. Ringbauer, C. J. Wood, K. Modi, A. Gilchrist,
A. G. White, and A. Fedrizzi, Phys. Rev. Lett. \textbf{114}, 090402
(2015).

\bibitem{Kuah07} A. M. Kuah, K. Modi, C. A. Rodr\'{i}guez-Rosario,
and E. C. G. Sudarshan, Phys. Rev. A \textbf{76}, 042113 (2007); K.
Modi and E. C. G. Sudarshan, Phys. Rev. A \textbf{81}, 052119 (2010).

\bibitem{Modi11} K. Modi, Open Syst. Inf. Dyn. \textbf{18}, 253 (2011). 

\bibitem{Gong13A} A. Z. Chaudhry and J. B. Gong, Phys. Rev. A \textbf{87},
012129 (2013). 

\bibitem{Gong13B} A. Z. Chaudhry and J. B. Gong, Phys. Rev. A \textbf{88},
052107 (2013).

\bibitem{Pechukas94}P. Pechukas, Phys. Rev. Lett. \textbf{73}, 1060
(1994). 

\bibitem{Royer96} A. Royer, Phys. Rev. Lett. \textbf{77}, 3272 (1996). 

\bibitem{Campisi09}M. Campisi, P. Talkner, and P. H\"anggi, Phys.
Rev. Lett. \textbf{102}, 210401 (2009). 

\bibitem{Dijkstra10}A. G. Dijkstra and Y. Tanimura, Phys. Rev. Lett.
\textbf{104}, 250401 (2010).

\bibitem{key-15}A. G. Dijkstra and Y. Tanimura, Phil. Trans. R. Soc.
A \textbf{370}, 3658 (2012).

\bibitem{Gong12}C. K. Lee, J. S. Cao, and J. B. Gong, Phys. Rev.
E \textbf{86}, 021109 (2012). 

\bibitem{key-01-2}H. A. Carteret, D. R. Terno, and K. \.{Z}yczkowski,
Phys. Rev. A \textbf{77}, 042113 (2008).

\bibitem{key-01-3}K. Modi, C. A. Rodr\'{i}guez-Rosario, and A. Aspuru-Guzik,
Phys. Rev. A \textbf{86}, 064102 (2012).

\bibitem{key-01-4}F. Buscemi, Phys. Rev. Lett. \textbf{113}, 140502
(2014).

\bibitem{key-01-5}L. Liu and D. M. Tong, Phys. Rev. A \textbf{90},
012305 (2014).

\bibitem{key-01-7}J. M. Dominy, A. Shabani, and D. A. Lidar, arXiv:
1312.0908v4.

\bibitem{Ban09}M. Ban, Phys. Rev. A \textbf{80}, 064103 (2009).

\bibitem{Uchiyama10} C. Uchiyama and M. Aihara, Phys. Rev. A \textbf{82},
044104 (2010). 

\bibitem{Smirne10}A. Smirne, H-P. Breuer, J. Piilo, and B. Vacchini,
Phys. Rev. A \textbf{82}, 062114 (2010). 

\bibitem{Zhang10}Y. J. Zhang, X-B. Zou, Y-J. Xia, and G-C. Guo, Phys.
Rev. A \textbf{82}, 022108 (2010).

\bibitem{Dajka11}J. Dajka and J. \L uczka, Phys. Rev. A \textbf{82},
012341 (2010); J. Dajka, J. \L uczka, and P. H\"anggi, ibid. \textbf{84},
032120 (2011). 

\bibitem{Morozov12}V. G. Morozov, S. Mathey, and G. R\"opke, Phys.
Rev. A \textbf{85}, 022101 (2012). 

\bibitem{Uchiyama12}C. Uchiyama, Phys. Rev. A \textbf{85}, 052104
(2012). 


\bibitem{Gao13}Y. Gao, Eur. Phys. J. D \textbf{67}, 183 (2013). 

\bibitem{Semin12}V. Semin, I. Sinayskiy, and F. Petruccione, Phys.
Rev. A \textbf{86}, 062114 (2012). 

\bibitem{key-14}E. M. Laine, J. Piilo, and H.-P. Breuer, Europhys.
Lett. \textbf{92}, 60010 (2010). 

\bibitem{key-24}C. A. Rodr\'{i}guez-Rosario, K. Modi, L. Mazzola,
and A. Aspuru-Guzik, Europhys. Lett. \textbf{99}, 20010 (2012). 

\bibitem{key-31}L. Mazzola, C. A. Rodr\'{i}guez-Rosario, K. Modi,
and M. Paternostro, Phys. Rev. A \textbf{86}, 010102(R) (2012). 

\bibitem{key-30}A. Smirne, L. Mazzola, M. Paternostro, and B. Vacchini,
Phys. Rev. A \textbf{87}, 052129 (2013).

\bibitem{Rivas14} \'{A}. Rivas, S. F. Huelga, and M. B. Plenio,
Rep. Prog. Phys. \textbf{77}, 094001 (2014) and references therein.

\bibitem{Gorini76}V. Gorini, A. Kossakowski, and E. C. G. Sudarshan,
J. Math. Phys. \textbf{17}, 821 (1976).

\bibitem{Lindblad76} G. Lindblad, Commun. Math. Phys. \textbf{48},
119 (1976).

\bibitem{Kurizki08}N. Erez, G. Gordon, M. Nest, and G. Kurizki, Nature
(London) \textbf{452}, 724 (2008).

\bibitem{Kurizki09} G. Gordon, G. Bensky, D. Gelbwaser-Klimovsky,
D. D. Rao, N. Erez, and G. Kurizki, New J. Phys. \textbf{11}, 123025
(2009). 

\bibitem{Kurizki10}G. Gordon, D. D. Rao, and G. Kurizki, New J. Phys.
\textbf{12}, 053033 (2010).

\bibitem{key-12}C. Meier and D. J. Tannor, J. Chem. Phys. \textbf{111},
3365 (1999).

\bibitem{key-26}X. Wang and S. G. Schirmer, Phys. Rev. A \textbf{79},
052326 (2009).

\bibitem{key-28}H.-P. Breuer, E.-M. Laine, and J. Piilo, Phys. Rev.
Lett. \textbf{103}, 210401 (2009).

\bibitem{key-20}R. X. Xu and Y. J. Yan, J. Chem. Phys. \textbf{116},
9196 (2002).

\bibitem{key-18} U. Kleinekath\"ofer, J. Chem. Phys. \textbf{121},
2505 (2004).

\bibitem{key-22}A. Pomyalov, C. Meier, and D. J. Tannor, Chem. Phys.
\textbf{370}, 98 (2010).

\bibitem{key-21}B. Hwang and H. S. Goan, Phys. Rev. A \textbf{85},
032321 (2012); J.-S. Tai, K.-T. Lin, and H.-S. Goan, Phys. Rev. A \textbf{89}, 062310
(2014).

\bibitem{key-23}E. Geva, E. Rosenman, and D. Tannor, J. Chem. Phys.
\textbf{113}, 1380 (2000).

\bibitem{key-25}C. H. Fleming, A. Roura, and B. L. Hu, Phys. Rev.
E \textbf{84}, 021106 (2011).






\end{thebibliography}
\end{document}